\documentclass[aps,prd,a4paper,showpacs,12pt]{article}
\usepackage[margin=1.0in]{geometry}
\usepackage{multicol}
\usepackage{flushend}
\usepackage[toc,page]{appendix}%%%%%%%%%%%%%%%%%%%%%%%%%%%%%%%%%%%%%%%%%%%%%%%%%%%%%%%%%%% New
\usepackage{subfigure}
\usepackage {graphicx}
\usepackage{color}
\usepackage{xcolor}
\usepackage{threeparttable}
\usepackage{amsfonts}
\usepackage{times}
\usepackage{setspace}
\usepackage{balance}
\usepackage{lastpage}
\usepackage{amsthm}
\usepackage[tbtags]{amsmath}
\usepackage{nomencl}
\usepackage{pstricks}
\usepackage{pstricks-add}
\usepackage{pst-plot,pstricks-add}

\usepackage{amsmath,amsfonts,amssymb,simplewick}
\usepackage{float}
\usepackage{amsmath,amssymb,amsfonts,epsfig,graphicx,euscript}%
\usepackage{hyperref} 
\hypersetup{bookmarks=true, bookmarksnumbered=true,linktoc=page, pdfstartview={FitH}, 
	colorlinks=true, 
	citecolor=red, 
	filecolor=blue, 
	linkcolor=blue, 
	urlcolor=blue}
\usepackage{amsmath}% http://ctan.org/pkg/amsmath
\usepackage{amsfonts}% http://ctan.org/pkg/amsfonts
\usepackage{graphicx}
\usepackage{caption}
\usepackage{float}
\usepackage[all]{hypcap}
\numberwithin{equation}{section}
\usepackage{cite}
\usepackage{calc}
\newlength{\spacer}
\newsavebox{\mybox}

\newcommand{\bse}{\begin{subequations}}
	\newcommand{\ese}{\end{subequations}}
\newcommand{\be}{\begin{equation}}
\newcommand{\ee}{\end{equation}}
\newcommand{\bea}{\begin{eqnarray}}
\newcommand{\eea}{\end{eqnarray}}
\newcommand{\ba}{\begin{array}}
	\newcommand{\ea}{\end{array}}

\renewcommand{\thefootnote}{\fnsymbol{footnote}}
\begin{document}
	
	%\hfill%
	%\vbox{
	%    \halign{#\hfil        \cr
	%           IPM/P-2012/010\cr
	%                     }}
	%\vspace{1cm}
	\begin{center}
		{ \large{\textbf{ Contribution of the thermal mass to the chiral vortical effect and magnetobaryogenesis}}} %\\
		%		{ \large{\textbf{ The contribution of the hypermagnetic field to the generation and evolution of the vorticity and the matter-antimatter asymmetry in the early Universe}}} %\\
		\vspace*{1.5cm}
		\begin{center}
			{\bf S. Abbaslu\footnote{s$_{-}$abbasluo@sbu.ac.ir}$^{1,2}$, S. Rostam Zadeh\footnote{sh$_{-}$rostamzadeh@ipm.ir}$^2$ and S. S. Gousheh\footnote{ss-gousheh@sbu.ac.ir}$^1$, }\\
			\vspace*{0.5cm}
			{\it{$^1$Department of Physics, Shahid Beheshti University, Tehran, Iran\\$^2$School of Particles and Accelerators, Institute for Research in Fundamental Sciences (IPM), P.O.Box 19395-5531, Tehran, Iran}}\\
			\vspace*{1cm}
		\end{center}
	\end{center}
	\begin{center}
		\today
	\end{center}
	%\vspace{.5cm}
	%\bigskip
	
	\renewcommand*{\thefootnote}{\arabic{footnote}}
	\setcounter{footnote}{0}

\begin{abstract} 
    We show that the chiral vortical effect can exist in a nonchiral electroweak plasma in thermal equilibrium using the effective thermal masses of the fermions in the symmetric phase. We use a nonperturbative formula for the vortical current, which has been recently obtained and is a functional of the dispersion relation. Then, taking into account the effect of fermion thermal mass in the dispersion relation, we show that the corresponding hyperelectric vortical current receives implicit thermal corrections proportional to $T^2$, from both the gauge and Yukawa interactions of the fermion. We show that the contributions of gauge interactions to the thermal masses in the total hyperelectric vortical current cancel out due to the gauge symmetries, while those of the Yukawa interactions do not cancel out and yield $\vec{J}_{\mathrm{cv}}\simeq(3g^{\prime}/256\pi^2)T^2\vec{\omega}$. We finally show that, due to this current, only small transient vorticity fluctuations about the zero background value in a nonchiral electroweak plasma in thermal equilibrium can activate the chiral vortical effect, leading to the generation of hypermagnetic fields and matter-antimatter asymmetries in the symmetric phase of the early Universe, in the temperature range $100\mbox{GeV} \le T\le 10\mbox{TeV}$, all starting from zero initial values, even in the presence of the weak sphaleron processes. 

 \end{abstract}

%%%%%%%%%%%%%%%%%%%%%%%%%%%%%%%%%%%%%%%%%%%%%%%%%%%%%%%%%%%%%%%%%%%%%%%%%%%%%
\section{ Introduction}

The large-scale magnetic fields \cite{1bm,2bm,1,2} and the baryon asymmetry in the Universe \cite{baryon1,baryon2,baryon3,baryon4,WMAP-2010qai} are two important puzzles in particle physics and cosmology, the origin and evolution of which have been investigated through various models \cite{c1,c2,c3,c4,c5,c6,b1}. The amplitude of the detected coherent magnetic fields in the intergalactic medium is of the order of $B\sim10^{-15}G$ \cite{magnetic1,magnetic2,magnetic3,magnetic4,magnetic5}, and the measured baryon
asymmetry of the Universe is of the order of $\eta_B\sim10^{-10}$ \cite{baryon3,WMAP-2010qai}.  
 In the symmetric phase of the early Universe before the electroweak phase transition (EWPT), these two seemingly unrelated problems are intertwined via the Abelian anomalous effects including the Abelian anomaly, $\nabla_{\mu}j^{\mu}\sim\vec{E}_{Y}.\vec{B}_{Y}$, and the chiral magnetic effect (CME) \cite{sd1,sd2,sd3,Fujita-2016igl,sh1,sh2,sh3,vb1,vh1,mas,ms1,kh1}. 
 
The anomalous transport effects, namely the chiral vortical effect (CVE) and the CME \cite{Vilenkin:1978hb,av1,avi1}, play important roles in particle physics and cosmology, particularly in the early Universe \cite{s1}. Generation of electric currents parallel to the vorticity and the magnetic field is generally referred to as the CVE and the CME, respectively. In the presence of these anomalous effects, the ordinary magnetohydrodynamic (MHD) equations are generalized to the equations of anomalous magnetohydrodynamics (AMHD) \cite{Giovannini-1997eg,Giovannini-2013oga}. In Ref.\ \cite{Giovannini-2016whv}, the evolution equations of AMHD are derived in the extreme relativistic regime and the vortical current has been discussed.
%
%\textcolor{magenta}{} \textcolor{green}{}
%\textcolor{red}{} \textcolor{green}{} \textcolor{green}{,} \textcolor{orange}{()} \textcolor{purple}{}
%

In a single-species chiral plasma of massless fermions, the chiral magnetic current in the Landau-Lifshitz frame is \cite{Neiman-2010zi,Landsteiner-2012kdw,Yamamoto16,Landsteiner-2016}
\begin{equation}\label{a-3}
\vec{J}_{\mathrm{cm},r}=rQ_{r}\left[\frac{Q_{r}\mu_{r}}{4\pi^2}\left(1-\frac{1}{2}\frac{\Delta n_{r}\mu_{r}}{\rho+p}\right)-\frac{1}{24}\frac{\Delta n_{r}T^{2}}{\rho+p}\right]\vec{B_{Y}},
\end{equation}
where $r=\pm 1$ denotes the chirality, $Q_{r}$ is the electric charge, $\Delta n_{r}$ is the  difference between the particle and anti-particle number densities, and $\mu_{r}$ is the chiral chemical potential. The latter two and the asymmetry $\eta_r$ are related by $\eta_r =\Delta n_r/s \simeq T^2\mu_r/(6s)$\footnote{We will state the exact relation between $\Delta n_r$ and $\mu_r$ in Sec.\ \ref{x3}, and justify the approximate relation mentioned here.}, where $s$ denotes the entropy density. Oftentimes in the literature $\Delta n_r$ is simply denoted by $n_r$. At high temperatures $\mu_{r}/T\ll1$, so the chiral magnetic current reduces to 
$\vec{J}_{\mathrm{cm},r}=\frac{rQ_{r}^2}{4\pi^2}\mu_{r}\vec{B_{Y}}$, which is indeed the chiral magnetic current in the Eckart frame. In studies which have considered the Abelian anomaly and the CME, an initially strong hypermagnetic field can generate  matter-antimatter asymmetries from zero initial values \cite{sh1,sh2,sh3,Abbaslu2,Temperature fluctuation}, or a large initial asymmetry can amplify a seed of the hypermagnetic field \cite{sh1,sh2,sh3,Abbaslu1,Abbaslu2}. In order for the CME to be successful in the latter scenario, there should be a hypermagnetic seed field, which can be produced via the CVE \cite{Abbaslu1}.
 
The evolution of the magnetic fields and matter asymmetries is also connected to that of the fluid vorticity through the CVE \cite{Tashiro-36,Giovannini-2015aea}.
It is shown that in a single-species chiral plasma of massless fermions, the chiral vortical current in the Landau-Lifshitz frame is \cite{son1,Neiman-2010zi,Landsteiner-2012kdw,Yamamoto16,Landsteiner-2016}
\begin{equation}
  \vec{J}_{\mathrm{cv},r}=rQ_{r}\left[\frac{\mu_{r}^{2}}{8\pi^2}\left(1-\frac{2}{3}\frac{\Delta n_{r}\mu_{r}}{\rho+p}\right)+\frac{1}{24}T^{2}\left(1-\frac{2\Delta n_{r}\mu_{r}}{\rho+p}\right)\right]\vec{\omega},
\end{equation}  
where $\vec{\omega}$ is the fluid vorticity.\footnote{There are also higher order terms of the vorticity, the lowest being $\omega^2 \vec{\omega}$. This term appears with opposite signs for chiral fermions, which cancel out in total vector currents and add up in the axial currents.}
Similar to the chiral magnetic current case, the chiral vortical current reduces to the form   
$\vec{J}_{\mathrm{cv},r}=rQ_{r}\left(\frac{T^2}{24}+\frac{\mu_{r}^2}{8\pi^2}\right)\vec{\omega}$ at high temperatures, which is the same as the one obtained in the Eckart frame.\footnote{In the Eckart frame, the chiral currents induce extra terms in the energy-momentum tensor \cite{Landsteiner-2012kdw,Gao-2017gfq}, while in the Landau-Lifshitz frame, the chiral currents acquire extra terms as mentioned above \cite{son1,Neiman-2010zi,Landsteiner-2012kdw,Yamamoto16,Landsteiner-2016}. Nevertheless, at high temperatures, these corrections in both frames are negligible. In this study, we use the Eckart frame.} Using this equation for all chiral fermions in thermal equilibrium in the symmetric phase, the explicit $T^2$ terms in the total hyperelectric vortical current cancel out due to the gauge symmetry \cite{Abbaslu1,Abbaslu2}. In the absence of these temperature-dependent terms, the CVE in a vortical and unbalanced chiral plasma can produce only an initial seed for the hypermagnetic field whose subsequent amplification needs another mechanism such as the CME \cite{Abbaslu1}. 
This seed can also be produced when there are temperature fluctuations for some matter degrees of freedom \cite{Temperature fluctuation}. So far most of the works in the literature on CME and CVE have concentrated on the massless fermionic limit, while very few works have discussed the CVE for the massive fermions \cite{Flachi-2017,Lin-12,Prokhorov-2020iln}. The chiral vortical current for a massive chiral fermion, in the small mass limit, is obtained as $\vec{J}_{\mathrm{cv},r}=rQ_{r}\left(\frac{T^2}{24}-\frac{m^2}{16\pi^2}+\frac{\mu_{r}^2}{8\pi^2}\right)\vec{\omega}$, where $m$ denotes the mass of the chiral fermion \cite{Flachi-2017,Lin-12,Prokhorov-2020iln}.

%
%\textcolor{magenta}{} \textcolor{green}{} \textcolor{green}{,} \textcolor{orange}{()} 
%\textcolor{olive}{} \textcolor{magneta}{} \textcolor{purple}{}
%

The source of the temperature-dependent part of the chiral vortical effect is still controversial, and the various approaches in this regard can be divided into two main categories. The first category is based upon statistical and thermodynamical approaches, which includes the Wigner function formalism in the kinetic theory \cite{Gao:2012ix,Gao:2018jsi} and the one based on the derivatives of the grand thermodynamic potential \cite{Kalaydzhyan}. The second approach utilizes the Kubo formula and the gravitational anomaly \cite{Land1,Landsteiner-2012kdw}. Emergence of this term within the first category has also been shown in various other cases such as free rotating fermions \cite{av1,avi1}, and anomalous chiral superfluids \cite{Volovik:2003fe, Basar:2013qia}. In a model which assumes that the origin of this term is due to the mixed gauge-gravity anomaly, the Yukawa interactions, contrary to the gauge interactions, do not seem to contribute to the radiative corrections\cite{Golkar:2012kb}. 

Recently a nonperturbative approach has been introduced to calculate the chiral vortical current using the covariant Wigner function for particles with spin $1/2$, as a functional of the dispersion relation, in accelerated and rotating media \cite{Prokhorov-2017atp,Prokhorov:2018qhq}. Moreover, a nonperturbative formula has been presented for the chiral vortical current of massive fermions as a functional of the Fermi-Dirac distribution function for the case of zero accelerartion, i.e., a = 0 \cite{Zakharov:2020ked}, in the same general framework. It has been shown that it exactly coincides with the predictions of the Zubarev density operator in the quantum statistical approach, in both cases of massless and massive fermions. The formulas presented in \cite{Prokhorov-2017atp,Prokhorov:2018qhq,Zakharov:2020ked} for the chiral vortical current are all functionals of the dispersion relation. As is well known, the finite temperature effect on the massless Dirac propagators can be approximated by shifting the pole, and hence the dispersion relation, and the resulting thermal masses have been calculated, {\it i.e.}, $m_{r}(T)=\lambda_{r}T$ \cite{Weldon-1982}\footnote{Although a chiral fermion is massless at the tree level in the symmetric phase, at temperature $T$ its dispersion relation becomes $E(k)\simeq\sqrt{{\vec k}^2+m_r(T)^2}$.}.
Taking into account the modified dispersion relation, as prescribed in Ref.\ \cite{Prokhorov-2020iln}, we show that the CVE receives thermal corrections from both the gauge and Yukawa interactions. The hyperelectric chiral vortical current for a chiral fermion, in the small thermal mass and chemical potential limit, is obtained as $\vec{J}_{\mathrm{cv},r}=rQ_{r}\left(\frac{T^2}{24}-\frac{m_r^2(T)}{16\pi^2}+\frac{\mu_{r}^2}{8\pi^2}\right)\vec{\omega}$.\footnote{In the broken phase, the contributions of fermion thermal masses to the total electric vortical current cancel out.} This correction term adds an implicit $T^2$ dependence.

%
%\textcolor{magenta}{} \textcolor{green}{} \textcolor{green}{,} \textcolor{orange}{()} 
%\textcolor{olive}{} \textcolor{magneta}{} \textcolor{purple}{}
%

 In this study, we include the effect of gauge and Yukawa interactions via thermal masses on the CVE, CME, pressure, energy density, number densities, and asymmetries. However, the effect is especially pronounced for the hyperelectric vortical current. As we shall show, the contributions of all thermal masses to the total CVE coefficient which originate from the gauge interactions cancel out due to the gauge invariance, while those of the Yukawa processes do not cancel out.  Therefore,  the CVE can become active in the presence of vorticity alone. This, as we shall show, can lead to the creation of a sufficiently strong hypermagnetic field, which then produces the particle asymmetries due to the Abelian anomaly, even when the weak sphalerons are present.

The organization of the paper is as follows: In Sec.\ \ref{x1}, we briefly review the effective thermal masses for the fermions. In Sec.\ \ref{x2}, we use the effective thermal masses for fermions, to obtain the total hyperelectric chiral vortical current in the symmetric phase. In Sec.\ \ref{x3}, we obtain the total hyperelectric chiral magnetic current, by taking into account the effective thermal masses for fermions.
 In Sec.\ \ref{x5}, we present the anomalous magnetohydrodynamic equations in the Eckart frame for the FRW metric. In Sec.\ \ref{equilibrium}, we state the equilibrium conditions in the temperature range $100\mbox{GeV} \le T\le 10\mbox{TeV}$. In Sec.\ \ref{x6}, we derive the complete set of evolution equations for the matter-antimatter asymmetries, and the hypermagnetic field, taking the CVE and the CME into account, and solve the evolution equations numerically. We then display and discuss the results. In Sec.\ \ref{conclusion}, we summarize our results and state our conclusions. 
%
%\textcolor{magenta}{} \textcolor{green}{}
%\textcolor{red}{} \textcolor{green}{} \textcolor{green}{,} \textcolor{orange}{()} \textcolor{purple}{}
%

\section{Thermal mass}\label{x1}

It is known that in the symmetric phase, $T> T_{EW}$, the $\rm SU_L(2)\times U_{Y}(1)$ symmetry is restored and the gauge bosons and fermions are all massless at tree level. However, at finite temperature the pole of the fermion propagator no longer remains at $k^2=0$, where $k^\mu$ is the fermion four-momentum \cite{Weldon-1982}. Indeed, the particle acquires an effective thermal mass $m_r$ of order $T$ and, to a very good approximation, its dispersion relation becomes $E(k)\simeq\sqrt{{\vec k}^2+m_r(T)^2}$ \cite{Weldon-1982}.\footnote{We should emphasize that the nature of effective thermal mass is different from that of the ordinary mass. It has been shown that no term of the form $\bar{\psi}\psi$ is induced in the effective Lagrangian, and hence the $\textrm{SU}_{\textrm{L}}(2)$ gauge symmetry is preserved in the symmetric phase \cite{Weldon-1982}.} This is due to the particle propagation in the early Universe plasma and its interaction with other particles in the background at temperature $T$. The effective thermal masses for different chiralities and families are  as follows \cite{Weldon-1982,Davidson-1994gn}:

\begin{equation}\label{eq3a}
\begin{split}
&m_{{e_i}_R}^{2}={g^{\prime}}^2\frac{T^2}{8}+h_{e_i}^{2}\frac{T^2}{8},\\&
m_{{e_i}_L}^{2}={g^{\prime}}^2\frac{T^2}{32}+3g^{2}\frac{T^{2}}{32}+h_{e_i}^{2}\frac{T^2}{16},\\&
m_{{\nu_i}_L}^{2}={g^{\prime}}^2\frac{T^2}{32}+3g^{2}\frac{T^{2}}{32},\\&
m_{{d_i}_R}^{2}={g^{\prime}}^2\frac{T^2}{72}+{g^{\prime\prime}}^{2}\frac{T^{2}}{6}+h_{d_i}^{2}\frac{T^2}{8},\\&
m_{{u_i}_R}^{2}={g^{\prime}}^2\frac{T^2}{18}+{g^{\prime\prime}}^{2}\frac{T^{2}}{6}+h_{u_i}^{2}\frac{T^2}{8},\\&
m_{{u_i}_L}^{2}=m_{{d_i}_L}^{2}\equiv m_{Q_i}^{2}={g^{\prime}}^2\frac{T^2}{288}+3g^{2}\frac{T^{2}}{32}+{g^{\prime\prime}}^{2}\frac{T^{2}}{6}+(h_{u_i}^2+h_{d_i}^{2})\frac{T^2}{16},
\end{split}
\end{equation} 
where, $m_{{e_i}_R}$ ($m_{{e_i}_L}$), $m_{{\nu_i}_L}$, $m_{{d_i}_R}$ ($m_{{u_i}_R}$), and $m_{Q_i}$ denote thermal masses of right-handed (left-handed) charged leptons, left-handed neutrinos, down (up) right-handed quarks, and left-handed quarks, respectively, and \lq{\textit{i}}\rq\ is the generation index. Furthermore, $g^{\prime\prime}$, $g^{\prime}$, and $g$ are the $\rm SU(3)$, $\rm U_{Y}(1)$, and $\rm SU_{L}(2)$ coupling constants, respectively, and  $h_f$s are the Yukawa coupling constants given as \footnote{The Yukawa Lagrangian density is given by $ \mathcal{L}_{Y}=\sum_{ij}\big[h_{l}^{ij}\bar{l}_{iL}\Phi e_{jR}+h_{u}^{ij}\bar{Q}_{i}\tilde{\Phi}u_{jR}+h_{d}^{ij}\bar{Q}_{i}\Phi d_{jR}\big]+h.c. $, where $\tilde{\Phi}=i\sigma_{2}\Phi$. The fermion bases are chosen such that for right-handed fermions, left-handed leptons and left-handed quarks, ${h_{f}}^{\dagger}h_{f}$,  $h_{l}{h_{l}}^{\dagger}$ and $h_{u}{h_{u}}^{\dagger}+h_{d}{h_{d}}^{\dagger}$ are diagonal, respectively. In this basis, $h^{\dagger}h$ and $hh^{\dagger}$ have the same eigenvalues equal to $2m_{f}^2/v^2$, where $v=246 \mbox{GeV}$ is the Higgs vacuum expectation value \cite{Davidson-1994gn,Fujita-2016igl,Kawamura-2018cpu}.}  %\cite{Fujita-2016igl,Kawamura-2018cpu}} 
\begin{equation}
\begin{split}
&h_{e_i}\doteqdot  \mathrm{diag}(2.94\times10^{-6},6.9\times10^{-4},1.03\times10^{-2}),\\&
h_{u_i}\doteqdot \mathrm{diag}(1.3\times10^{-5},7.3\times10^{-3},1.0),\\&
h_{d_i}\doteqdot \mathrm{diag} (2.7\times10^{-5},5.5\times10^{-4},2.4\times10^{-2}).
\end{split}
\end{equation}
Upon substituting the relevant coupling constants in Eqs.\ (\ref{eq3a}), we obtain the effective thermal masses for the fermions as given in Table \ref{tableai}. %The effective thermal masses influence the physical quantities and anomalous transport effects, most importantly the Chiral Vortical Effect (CVE). 
Although thermal masses influence various physical quantities, we shall show that their most significant effect in this study is that in the presence of vorticity they become a new source for producing the hyperelectric vortical current.
%
%\textcolor{magenta}{} \textcolor{green}{}
%\textcolor{red}{} \textcolor{green}{} \textcolor{green}{,} \textcolor{orange}{()} \textcolor{purple}{}
%
\begin{table}[ht]
	\centering 
	\begin{center}
		\begin{tabular}{|p{40mm}|c|c|c|c| c| c|} 
			\hline
			\footnotesize {particle} &$\frac{m_r}{T}$ \\[0.5ex] 
			\hline 
			$e_{R}$, $\mu_R$, $\tau_R$ & $0.1220$  \\
			\hline
			$e_L$, $\mu_L$, $\tau_L$, $\nu_{e_L}$, $\nu_{\mu_L}$, $\nu_{\tau_L}$&$0.2027$ \\
			\hline
			$q_{d_R}$, $q_{s_R}$, $q_{b_R}$, $q_{u_R}$, $q_{c_R}$& $0.4817$ \\  
			\hline
			$q_{d_L}$, $q_{s_L}$, $q_{u_L}$, $q_{c_L}$&$0.5178$\\
			\hline 
			$q_{b_L}$, $q_{t_L}$&$0.5750$\\  
			\hline
			$q_{t_R}$&$0.5975$ \\[1ex] 
			\hline
		\end{tabular}
	\end{center}
	\caption{The numerical values of the effective thermal mass $m_{r}$ for all chiral fermions.}\label{tableai} 
\end{table}
%creation and evolution of magnetic fields and matter-antimatter asymmetries in the Universe \cite{s1}.
%
%\textcolor{red}{} \textcolor{green}{} \textcolor{green}{,} \textcolor{orange}{()} \textcolor{purple}{}
%

\section{CVE originating from fermion thermal masses}\label{x2}
The chiral vortical effect, generically induced by the rotation of chiral matter, refers to the generation of an electric current parallel to the vorticity field \cite{av1,Vilenkin:1978hb}. As mentioned in the Introduction, we use a nonperturbative formula for the  chiral vortical current, which has been obtained using the covariant Wigner function for particles with spin 1/2, as presented in \cite{Prokhorov-2020iln,Prokhorov:2018qhq}:
\begin{equation}\label{cvA1}
	\begin{split}
		j_{\mathrm{cv}}^{\mu 5}=&\frac{1}{2}\int \frac{d^{3}k}{(2\pi)^3}\left[f_{FD}\left(E_k -\mu-\frac{|\omega|}{2}\right)-f_{FD}\left(E_k-\mu+\frac{|\omega|}{2}\right)\right. \\&
	\left.	+f_{FD}\left( E_k +\mu-\frac{|\omega|}{2} \right) 
	-f_{FD}\left( E_k+\mu+\frac{|\omega|}{2} \right)\right]\frac{\omega^{\mu}}{|\omega|},
	\end{split}
\end{equation}
where $E_{k}=\sqrt{\vec k^2+m^2}$, $f_{FD}(x)=\frac{1}{e^{\beta x}+1}$ is the Fermi-Dirac distribution function\footnote{To a good approximation, when departures from thermal equilibrium are moderate and the interaction rates are high compared to the Hubble rate, the distribution function of a given species is well described by the equilibrium distribution function \cite{Escudero,Kolb}.}, $\beta=1/T$, and $m$ and $\mu$ are the mass and chemical potential of the fermion, respectively. Furthermore, $\omega^{\mu}=(\epsilon^{\mu\nu\rho\sigma}/{R(t)}^{3})u_{\nu}\nabla_{\rho}u_{\sigma}$ is the vorticity four-vector, with the totally anti-symmetric four-dimensional Levi-Civita symbol specified by $\epsilon^{0123}=-\epsilon_{0123}=1$, $u^{\mu}=\gamma\left(1,\vec{v}/R(t)\right)$ is the four-velocity of the plasma normalized such that $u^{\mu}u_{\mu}=1$, $\gamma$ is the Lorentz factor, $R(t)$ is the scale factor, and $\nabla_{\mu}$ is the covariant derivative with respect to the Friedmann-Robertson-Walker (FRW) metric, $ds^{2}=dt^{2}-R^{2}(t)\delta_{ij}dx^{i}dx^{j}$.
After expanding in terms of $\omega$ up to the linear term, Eq.\ (\ref{cvA1}) simplifies to \cite{Flachi-2017,Lin-12}
\begin{equation}\label{eqxiv}
		j_{\mathrm{cv}}^{\mu 5}=-\int_{0}^{\infty} \frac{dk}{4\pi^2}k^2\left[\frac{d}{dE_k}\tilde{F}_{+}(E_{k},\mu)\right]\omega^{\mu} =\frac{1}{4\pi^2}\int_{0}^{\infty}dk\left[\frac{k^2+{E_{k}}^2}{E_{k}}\right]\tilde{F}_{+}(E_{k},\mu)\omega^{\mu},
\end{equation}
where $\tilde{F}_{+}(E_{k},\mu)=f_{FD}(E_{k}-\mu)+f_{FD}(E_{k}+\mu)$. For the case $\mu=0$, this expression simplifies to \cite{Prokhorov-2020iln},
\begin{equation}\label{cvA2}
	\begin{split}
			j_{\mathrm{cv}}^{\mu 5}=\frac{T^2}{2\pi^2}\int_{0}^{\infty}dx \left[\frac{x^2 e^{\sqrt{x^2+(\frac{m}{T})^2}}}{(1+e^{\sqrt{x^2+(\frac{m}{T})^2}})^2}\right]\omega^{\mu} =\frac{1}{2}T^2 c\left(\frac{m}{T}\right)\omega^{\mu},
	\end{split}
\end{equation}
%
%\textcolor{magenta}{} \textcolor{green}{} \textcolor{green}{, \it i.e.},
%\textcolor{red}{} \textcolor{green}{} \textcolor{green}{,} \textcolor{orange}{()} \textcolor{purple}{}
%
where $x=k/T$.\footnote{The presence of the additional factor of $\frac{1}{2}$ in front of Eqs.\ (\ref{cvA1}) and (\ref{cvA2}) as compared to Ref.\ \cite{Prokhorov-2020iln} is due to the difference between our definitions of $\omega^{\mu}$.} When the effective thermal mass of particle $m_{eff}=\lambda T$ is used in the dispersion relation, the coefficient $c(\frac{m}{T})$ becomes independent of temperature and less than $1/6$,  {\it i.e.}, $c(\frac{m}{T}) \le c(0)=\frac{1}{6}$ \cite{Prokhorov-2020iln}. It is precisely this fact that we use in this study. That is, we use thermal mass in the dispersion relation whenever it appears in equations such as Eqs.\ (\ref{eqxiv}, \ref{cvA2}).\footnote{We do the same when we calculate the chiral magnetic current, the number density, the energy density, the pressure, and the asymmetry number density.}

The hyperelectric chiral vortical current for one fermion species with two different handedness is defined as 
\begin{equation}\label{CVEA1}
		{J}^{\mu}_{{\mathrm{cv}},r} := \frac{Q_{r}}{2}[j_{\mathrm{cv}}^{\mu }+r	j_{\mathrm{cv}}^{\mu 5}]=:Q_{r}\xi_{\mathrm{v,}r}(T,m_r,\mu_r)\omega^{\mu} =:c_{{\mathrm{v},r}}\omega^{\mu},
\end{equation}
where $r = \pm1$ represents the chirality, $Q_r=-g^{\prime}Y_{r}/2$, $Y_{r}$ is the hypercharge of the chiral fermion, and $c_{{\mathrm{v},r}}$ is the hyperelectric chiral vortical coefficient. Since $j_{\mathrm{cv}}^{\mu }=0$ \cite{Becattini-2013fla}, the expression for $\xi_{\mathrm{v,}r}(T,m_r,\mu_r)$ can be inferred from  Eqs.\ (\ref{eqxiv}, \ref{CVEA1}):
\begin{equation}\label{eqxiv2}
	\xi_{\mathrm{v,}r}(T,m_r,\mu_r)=\frac{r}{8\pi^2}\int_{0}^{\infty}dk\left[\frac{k^2+{E_{k}}^2}{E_{k}}\right]\tilde{F}_{+}(E_{k},\mu),
\end{equation}

In the symmetric phase of the early Universe plasma at high temperatures, $\mu_r/T\ll1$\footnote{The maximum asymmetry is $\eta \simeq 10^{-10}$, for which $\mu/T \simeq 10^{-7}$.}. Using this fact, along with $m_r/T<1$, as shown in Table \ref{tableai}, we can Taylor expand the expression for $\xi_{\mathrm{v,}r}$ given by Eq.\ (\ref{eqxiv2}) to obtain (see Appendix \ref{Appendix-A})\footnote{In Appendix \ref{Appendix-B} we show that although the values of $m_r/T$ are not much less than one, the small thermal mass approximation yields unexpectedly good results. To be specific, we first neglect the tiny $\mu_r$ terms, then show that the results based on the small thermal mass approximation, given in Eq.\ (\ref{eqxiv3}), and the results obtained by numerical integration of the exact expression, given in Eq.\ (\ref{eqxiv2}), are within $0.25\%$.}  \cite{Flachi-2017,Lin-12,Prokhorov-2020iln}
\begin{equation}\label{eqxiv3}
\xi_{\mathrm{v,}r}(T,m_{r},\mu_r)\simeq r\left(\frac{T^2}{24}-\frac{m_{r}^2}{16\pi^{2}}+\frac{\mu_{r}^2}{8\pi^2}\right).
\end{equation}
For a massless fermion with zero thermal mass this reduces to the form $\xi_{\mathrm{v,}r}(T,0,\mu_r)=r\left( \frac{T^2}{24}+\frac{\mu_{r}^2}{8\pi^2}\right)$ \cite{av1,son1,Lin-12,a1,a11,a2,a3,a4,a5,a6}. 
%\textcolor{red}{$m_r/T\ll1$} \textcolor{blue}{$m_r/T<1$}

Using Eqs.\ (\ref{CVEA1},\ref{eqxiv3}) for all chiral fermions in the plasma, the hyperelectric chiral vortical coefficient $c_{\mathrm{v}}$ can be obtained as 
\begin{equation}\label{eq222}
	\begin{split}
	c_{\mathrm{v}}(t)&=\sum_{i=1}^{n_{G}}\Big[\frac{g'}{48}\Big(-Y_{e_{R}}T_{{e_i}_R}^{2}+Y_{e_{L}}T_{{e_i}_L}^{2}+Y_{e_{L}}T_{{\nu_i}_L}^{2}-Y_{d_{R}}T_{{d_i}_R}^{2}N_{c}-Y_{u_{R}}T_{{u_i}_R}^{2}N_{c}+Y_{Q}T_{u_{i_L}}^{2}N_{c}+Y_{Q}T_{{d_i}_L}^{2}N_c\Big)\\&-\frac{{g'}}{32\pi^{2}}\Big(-Y_{e_{R}}m_{{e_i}_R}^{2}+Y_{e_{L}}m_{{e_i}_L}^{2}+Y_{e_{L}}m_{{\nu_i}_L}^{2}-Y_{d_{R}}m_{{d_i}_R}^{2}N_{c}-Y_{u_{R}}m_{{u_i}_R}^{2}N_{c}+Y_{Q}m_{{u_i}_L}^{2}N_{c}+Y_{Q}m_{{d_i}_L}^{2}N_{c}\Big)\\&+\frac{{g'}}{16\pi^{2}}\Big(-Y_{e_R}\mu_{{e_i}_R}^{2}+Y_{e_L}\mu_{{e_i}_L}^{2}+Y_{e_L}\mu_{{\nu_i}_L}^{2}-Y_{d_{R}}\mu_{{d_i}_R}^{2}N_{c}-Y_{u_{R}}\mu_{{u_i}_R}^{2}N_{c}+Y_{Q}\mu_{{u_i}_L}^{2}N_{c}+Y_{Q}\mu_{{d_i}_L}^{2}N_{c}\Big)\Big], 
	\end{split}
\end{equation}
%
%\textcolor{magenta}{} \textcolor{green}{} \textcolor{green}{, \it i.e.},
%\textcolor{red}{} \textcolor{green}{} \textcolor{green}{,} \textcolor{orange}{()} \textcolor{purple}{}
%
where, $\mu_{{e_i}_R}$ ($\mu_{{e_i}_L}$), $\mu_{{\nu_i}_L}$ ,  $\mu_{{u_i}_R}$ ($\mu_{{u_i}_L}$), and  $\mu_{{d_i}_R}$ ($\mu_{{d_i}_L}$) denote the chemical potential of right-handed (left-handed) charged leptons, left-handed neutrinos,  right-handed (left-handed) up quarks, and right-handed (left-handed) down quarks, respectively.
In the above equation, $n_{G}$ is the number of generations, and $N_{c}=3$ is the rank of the non-Abelian $\rm SU(3)$ gauge group. The relevant hypercharges are 
\begin{equation}\label{eqds2}
	\begin{split}
		Y_{e_{L}}=-1,\ \ Y_{e_{R}}=-2,\ \ Y_{Q}=\frac{1}{3},\ \ Y_{u_{R}}=\frac{4}{3},\ \ Y_{d_{R}}=-\frac{2}{3}.
	\end{split}	
\end{equation}
For temperatures $T\le 10^{14} \mbox{GeV}$, the Abelian $\rm U_Y(1)$ gauge interactions cause all particles to be in thermal equilibrium, and therefore have the same temperature \cite{Bodeker-2019}. Moreover, rapid non-Abelian $\rm SU_{L}(2)$ and $\rm{ SU(3)}$ gauge interactions enforce the equality of the number density carried by different components of a given $\rm SU_L(2)$ and $\rm SU(3)$ multiplet, e.g.\ $\mu_{e_{iL}}=\mu_{\nu_{iL}}$ and $\mu_{u_{iL}}=\mu_{d_{iL}}\equiv\mu_{Q_i}$, regardless of the color. In fact, color-independence has already been assumed in writing Eq.\ (\ref{eq222}). 

As mentioned earlier, all chiral fermions acquire thermal masses, given by Eq.\ (\ref{eq3a}). When we substitute the fermion thermal masses and the hypercharges, given by Eqs.\ (\ref{eq3a},\ref{eqds2}), into Eq.\ (\ref{eq222}), two major cancellations occur due to gauge symmetry. First, the original and explicit $T^2$ terms which appear in the first parenthesis cancel each other. Second, all terms coming from the contributions of gauge interactions to thermal masses in the second parenthesis of Eq.\ (\ref{eq222}) cancel one another exactly, as well. Therefore, only the contributions due to the Higgs coupling to thermal masses remain and Eq.\ (\ref{eq222}) simplifies to
\begin{equation}\label{eq321ab2}
	\begin{split}	c_{\mathrm{v}}(t)=&\sum_{i=1}^{n_G}\Big[-\frac{g^{\prime}}{32\pi^{2}}\Big(3h_{e_i}^{2}\frac{T^2}{16}+3h_{d_i}^{2}\frac{T^2}{8}-3h_{u_i}^{2}\frac{T^2}{8}\Big)\\&+\frac{{g'}}{8\pi^{2}}\Big(\mu_{{e_i}_R}^{2}-\mu_{{e_i}_L}^{2}+\mu_{{d_i}_R}^{2}-2\mu_{{u_i}_R}^{2}+\mu_{Q_{i}}^{2}\Big)\Big].	
	\end{split}
\end{equation}

We should emphasize that the original and explicit $T^2$ dependence in the expression for $c_{\mathrm{v,}r}$ given in Eq.\ (\ref{eq222}), i.e.\ $-r\frac{g'}{48}Y_{r}T^{2}$, which added up to zero, is distinct from the $T^2$ dependence which entered Eq.\ (\ref{eq321ab2}) implicitly from the Higgs coupling component of thermal masses $m_r(T)$, i.e.\ $r\frac{g^{\prime}}{32\pi^{2}}Y_r m^{2}_r(T)$ which is $\frac{g^{\prime}}{32\pi^{2}}Y h^{2}\frac{T^2}{8}$ for the right-handed particles and $-\frac{g^{\prime}}{32\pi^{2}}Y h^{2}\frac{T^2}{16}$ for the left-handed ones.
When the thermal masses are neglected, Eqs.\ (\ref{eq222},\ref{eq321ab2}) reduce to the usual forms \cite{Abbaslu1}, and therefore there would be no CVE in a chirally balanced plasma. However,  as Eqs.\ (\ref{eq222},\ref{eq321ab2}) indicate, even in the absence of chirality, $c_{\mathrm{v}}$ is nonzero due to thermal masses. Hence, the activation of CVE in the symmetric phase only requires the presence of a vorticity $\omega^{\mu}$.

Let us neglect the tiny $\mu_r^2$ terms in Eq.\ (\ref{eq321ab2}) and focus on the role of thermal masses in the CVE. Since the top quark Yukawa coupling $h_t$ is very much larger than other Yukawa couplings, it suffices to consider only $h_t$ and neglect all other $h_f$s and obtain\footnote{Using Eq.\ (\ref{CVEA1}) and the numerical values  $\xi_{\mathrm{v,r,N}}(T,m_r)$ given in Table \ref{tablea} for all particles, the exact value is $c_{\mathrm{v,N}}(T)=(0.00096)g^{\prime}T^2$.}
%Using Eqs.\ (\ref{CVEA1},\ref{eqxiv}) and effective thermal masses given in Table \ref{tablea}, $c_\mathrm{v}$ can be obtained numerically as $c_{\mathrm{v,N}}(t)=(0.001)g^{\prime}T^2$, which is equal to $c_\mathrm{v}$ obtained in Eq.\ (\ref{eq321}), with a very good accuracy.}
%which after substituting its value we obtain the $c_\mathrm{v}$ as
\begin{equation}\label{eq321}
\begin{split}
c_{\mathrm{v}}(t)\simeq\frac{3g^{\prime}}{256\pi^{2}}T^2= (0.00118)g^{\prime}T^2.
\end{split}
\end{equation}
%
%\textcolor{red}{} \textcolor{green}{} \textcolor{green}{,} \textcolor{orange}{()} \textcolor{purple}{}
%
As can be seen, there exists a significant contribution to the hyperelectric vortical current which might be consequential at high temperatures in the symmetric phase.\footnote{It is worth mentioning that although after the EWPT fermions acquire bare masses, the contributions of these masses to the electric chiral vortical current cancel out, since the right- and left-handed fermions have the same masses and electric charges.} 
This is the first main result of this study. As we shall show,  CVE coming from fermion thermal masses plays an important role in the generation of the hypermagnetic fields and the matter-antimatter asymmetries.
%Let us now obtain the hyperelectric chiral vortical coefficient $c_\mathrm{v}$ numerically. Using the Eqs.\ (\ref{CVEA1}), (\ref{CVE1}) and given effective thermal mass from Table \ref{tablea} we obtain the $c_\mathrm{v}$ numerically as 
%\begin{equation}
%c_{\mathrm{v,N}}(t)=(0.001)g^{\prime}T^2.
%\end{equation}
%As can be see the numerical value of the $c_\mathrm{v,N}$ with a good accuracy is equal to the obtained $c_\mathrm{v}$ at small mass limit. 

\section{Chiral magnetic effect } \label{x3}
The chiral magnetic effect refers to the generation of an electric current parallel to the magnetic field in an imbalanced chiral plasma. The hyperelectric chiral magnetic current for one fermion species with two different handedness is
%In an imbalanced chiral plasma, there exists an electric current parallel to the magnetic field. This phenomenon is called the Chiral Magnetic Effect (CME). The hyperelectric chiral magnetic current is given as \cite{,Lin-12} 
 \begin{equation}\label{qsh}
J^{\mu}_{\mathrm{B,}r}=Q_{r}\xi_{\mathrm{B,}r}(T,m_r,\mu_{r})B^{\mu} =c_{\mathrm{B,}r}B^{\mu}. 
\end{equation}
 Here, $B^{\mu}=(\epsilon^{\mu\nu\rho\sigma}/2R(t)^3)u_{\nu}Y_{\rho\sigma}={\tilde{Y}}^{\mu\nu}u_{\nu}$ is the hypermagnetic field four-vector, $Y_{\alpha\beta}=\nabla_{\alpha}Y_{\beta}-\nabla_{\beta}Y_{\alpha}$ is the field strength tensor of the $\rm U_{Y}(1)$ gauge fields, $c_{{\rm B},r}$ is the hyperelectric chiral magnetic coefficient, and similarly to $\xi_{\mathrm{v},r}(T,m_r,\mu_{r})$, $\xi_{\mathrm{B,}r}(T,m_r,\mu_{r})$ can be obtained as \cite{Lin-12,Gao-2020pfu}
 \begin{equation}\label{q2}
 \xi_{\mathrm{B,}r}(T,m_r,\mu_r)=\frac{rQ_{r}}{4\pi^2}\int_{0}^{\infty}dk\tilde{F}_{-}(E_{k,r},\mu_r),
 \end{equation}
 where $\tilde{F}_{-}(E_{k,r},\mu_r)=f_{FD}(E_{k,r}-\mu_{r})-f_{FD}(E_{k,r}+\mu_{r})$. 
    
Before the EWPT,  $\mu_r/T\ll1$  and the terms of order $\mu_r/T$ are usually neglected in the expressions for physical quantities such as the  number densities, energy density, pressure, entropy density, etc. However, for having nonzero CME, the Fermi-Dirac distribution function should be expanded to the first order in ($\mu_r/T$). In fact, in the $\mu_r=0$ limit, the C-symmetry requires $\xi_{\mathrm{B},r}=0$, so the interdependent CME and Abelian anomaly effects disappear.  A Taylor expansion of $\tilde{F}_{-}(E_{k,r},\mu_r)$ in Eq.\ (\ref{q2}), in terms of $(\mu_r/T)$ and $(m_r/T)$ yields (see Appendix \ref{Appendix-C})
\begin{equation} \label{Bsh02}
	\xi_{\mathrm{B},r}(T,m_r,\mu_r)=\frac{rQ_{r}\mu_r}{4\pi^2}\left[ 1+7\zeta^{'}(-2)\left(\frac{m_{r}}{T}\right)^2+\frac{31}{6} \zeta^{'}(-4)\left(\frac{m_{r}}{T}\right)^2\left(\frac{\mu_{r}}{T}\right)^2+\dotsb\right],
\end{equation}
 where $\zeta$ is Riemann zeta function and prime denotes the derivative, with $\zeta^{'}(-2)=-0.030448$ and $\zeta^{'}(-4)=0.00798381$.
Assuming $m_r=0$ and neglecting terms of higher order in $(\mu_r/T)$, this expression reduces to the simple form $\xi_{\mathrm{B},r}(T,0,\mu_r)=\frac{rQ_{r}}{4\pi^2}\mu_r$ 
	\cite{avi1,khar1,khar2,khar3}.

Since there is a prefactor $\mu_r$ in Eq.\ (\ref{Bsh02}), we can neglect the correction terms in the square bracket which are $\mathcal{O}[(m_{r}/T)^2]$ and $\mathcal{O}[(\mu_{r}/T)^2]$.\footnote{In Appendix \ref{Appendix-C}, we also show explicitly that the effects of thermal masses on the hyperelectric chiral magnetic coefficient $\xi_{\mathrm{B},r}$, are small. Moreover, in Appendix \ref{Appendix-D}, we show that the effects of thermal masses on four important physical quantities, namely, the number density $n_{r}$, the energy density $\rho_{r}$, the pressure $p_{r}$, and the asymmetry number density $\Delta n_{r}=(n_{r}-\bar{n}_{r})$, are small. In particular we justify the relation $\Delta n_r \simeq T^2\mu_r/6$ mentioned in the Introduction.} Hence, using Eq.\ (\ref{qsh}), the total hyperelectric chiral magnetic coefficient $c_\mathrm{B}$ can be written as
%  Considering all chiral fermions and neglecting $\delta(\xi_\mathrm{B},m_i)$ (see Appendix B), the total hyperelectric chiral magnetic coefficient $c_\mathrm{B}$ can be obtained as
 \begin{equation}\label{eqc_B} 
	\begin{split}
	c_{B}(t)=& 
	\frac{g'^{2}}{16\pi^{2}}\sum_{i=1}^{n_{G}}\Big[Y_{e_{R}}^{2}\mu_{{e_i}_R}-Y_{e_{L}}^{2}\mu_{{e_i}_L}N_{w}+Y_{d_{R}}^{2}\mu_{{d_i}_R}N_{c}\\&+Y_{u_{R}}^{2}\mu_{{u_i}_R}N_{c}-Y_{Q}^{2}\mu_{Q_{i}}N_{c}N_{w}\Big],
	\end{split}
	\end{equation}
where $N_{w}=2$ is the rank of the non-Abelian $\rm SU_{L}(2)$ gauge group.
In Sec.\ \ref{equilibrium} we will simplify the total hyperelectric chiral magnetic coefficient $c_\mathrm{B}$ by considering the equilibrium conditions in the plasma.

%
%\textcolor{magenta}{} \textcolor{green}{} \textcolor{green}{, \it i.e.},
%\textcolor{red}{} \textcolor{green}{} \textcolor{green}{,} \textcolor{orange}{()} \textcolor{purple}{}
%

%
%\textcolor{magenta}{} \textcolor{green}{} \textcolor{green}{, \it i.e.},
%\textcolor{red}{} \textcolor{green}{} \textcolor{green}{,} \textcolor{orange}{()} \textcolor{purple}{}
%

\section{Anomalous magnetohydrodynamics}\label{x5}
The equations of AMHD consist of the anomalous energy-momentum conservation and Maxwell's equations.
The energy-momentum conservation can be expressed covariantly as \cite{son1,Yamamoto16,Anand-2017,Landsteiner-2016}
\begin{equation}\label{consT}
\nabla_{\mu}T^{\mu\nu}= 0,
\end{equation}
where $T^{\mu\nu}$ is the energy-momentum tensor. In the presence of the CME and the CVE, the energy-momentum tensor acquires the following extra contributions in the Eckart frame \cite{Landsteiner-2016,Gao-2017gfq}
\begin{equation}
T^{\mu\nu}(CME)=c_\mathrm{v}\big[u^{\mu}B^{\nu}+u^{\nu}B^{\mu}\big],
\end{equation}
and
\begin{equation}
T^{\mu\nu}(CVE)=\frac{1}{2}\Delta n_{5,t}\big[u^{\mu}\omega^{\nu}+u^{\nu}\omega^{\mu}\big],
\end{equation}
where $\Delta n_{5,t}$ is the total axial asymmetry density given by
\begin{equation}
\begin{split}
\Delta n_{5,t}=&\sum_{i=1}^{n_G}\big[\Delta n_{{e_i}_R}-N_{w}\Delta n_{{e_i}_L}+N_{c} \Delta n_{{d_i}_R}+N_{c}\Delta n_{{u_i}_R}-N_{w}N_{c}\Delta n_{Q_i}\big].
\end{split}
\end{equation}

Therefore, the energy-momentum tensor $T^{\mu\nu}$ and the total hyperelectric current $J^{\mu}$ in the presence of the CME and the CVE for the early Universe plasma are given as  %\cite{son1,Yamamoto16,Anand:2017,Landsteiner-2016} 
%
%\textcolor{magenta}{} \textcolor{green}{} \textcolor{green}{, \it i.e.},
%\textcolor{red}{} \textcolor{green}{} \textcolor{green}{,} \textcolor{orange}{()} \textcolor{purple}{}
%
\begin{equation}
\begin{split}
T^{\mu\nu}&=(\rho+p)u^{\mu}u^{\nu}- p g^{\mu\nu}+\frac{1}{4}g^{\mu\nu} Y^{\alpha\beta} Y_{\alpha\beta}-Y^{\nu\sigma}{Y^{\mu}}_{\sigma} \\&+\tau^{\mu\nu}+c_\mathrm{v}\big[u^{\mu}B^{\nu}+u^{\nu}B^{\mu}\big]+\frac{1}{2}\Delta n_{5,t}\big[u^{\mu}\omega^{\nu}+u^{\nu}\omega^{\mu}\big]
,
\end{split}
\end{equation}
\begin{equation}\label{eq-ab22}
J^{\mu}=\rho_{\mathrm{el}} u^{\mu}+J^{\mu}_\mathrm{cm}+J^{\mu}_\mathrm{cv}+\nu^{\mu},
\end{equation}
where $\rho=g^{*}\pi^{2}T^{4}/30$ is the energy density, $p=\frac{1}{3}\rho$ is the pressure, $\rho_{\mathrm{el}}$ is the hyperelectric charge density, and $\nu^{\mu}\simeq \sigma E^{\mu}$ and $\tau^{\mu\nu}$ are the hyperelectric dissipative current and viscous stress tensor, respectively \cite{son1}.\footnote{The chiral particle dissipative current for a single species is $\nu^{\mu}_{r}=\sigma_{r} [E^{\mu}+T(u^{\mu}u^{\nu}-g^{\mu\nu})\nabla_{\nu}\big(\mu_r/T\big)]$, where  $\sigma_{r}\sim T$ is the relevant chiral conductivity \cite{son1,Yamamoto16,Anand-2017,Landsteiner-2016}. Since $\mu_r/T\ll1$, we will consider only the Ohmic part of the dissipative current, $\nu_{r}^{\mu}$, and neglect the second part \cite{Anand-2017}. Due to the homogeneity of the early Universe plasma, we can also neglect the gradient of pressure, temperature, and chemical potential in our AMHD equations.}  

Hereafter, we focus on high temperature and small velocity approximation, $v\ll1$. Then, the expressions for the four-vectors $B^{\mu}, \omega^{\mu}$, and $E^{\mu}$ become\footnote{In the derivative expansion of the hydrodynamics, $\partial_{t}\sim\vec{\nabla}.\vec{v}$, so  $\vec{v}\times\vec{E}_{Y}\simeq v^{2}\vec{B}_{Y}$, and we ignore the terms of $O(v^2)$ \cite{Yamamoto16}.} 
%
%\textcolor{red}{} \textcolor{green}{} \textcolor{green}{,} \textcolor{orange}{()} \textcolor{purple}{}
%
\begin{equation}\label{eq38}
	\begin{split}
		&B^{\mu}=\gamma\left(\vec{v}.\vec{B}_{Y},\frac{\vec{B}_{Y}-\vec{v}\times\vec{E}_{Y}}{R}\right)\simeq\left(\vec{v}. \vec{B}_{Y},\frac{\vec{B}_{Y}}{R}\right),\\&
		\omega^{\mu}=\gamma\left(\vec{v}.\vec{\omega},\frac{\vec{\omega}-\vec{v}\times\vec{a}}{R}\right)
		\simeq	\left(\vec{v}.\vec{\omega},\frac{\vec{\omega}}{R}\right),\\&
			%\textcolor{red}{a^\mu=\gamma\left(\vec{v}.\vec{a},\frac{\vec{a}+\vec{v}\times\vec{\omega}}{R}\right)\simeq \left(\vec{v}.\vec{a},\frac{\vec{a}}{R}\right),\\&}
		E^{\mu}=\gamma\left(\vec{v}.\vec{E}_{Y},\frac{\vec{E}_{Y}+\vec{v}\times\vec{B}_{Y}}{R}\right) \simeq \left(\vec{v}.\vec{E}_{Y},\frac{\vec{E}_{Y}+\vec{v}\times\vec{B}_{Y}}{R}\right).
		\end{split}
\end{equation}
%\textcolor{red}{where $a^{i}=R \Omega^{0i}$ is the acceleration three-vector, and $\Omega_{\mu\nu}=\nabla_{\mu}u_\nu-\nabla_{\nu}u_\mu$ is the vorticity tensor, }
Using Eqs.\ (\ref{eq-ab22}) and (\ref{eq38}), the temporal and spatial components of the total hyperelectric current $J^{\mu}=\left(\rho_{\mathrm{total}},\vec{J}/R\right)$ become 
	\begin{equation}\label{eq59-6}
		\begin{split}
			&\vec{J}=\rho_{el}\vec{v}+c_\mathrm{B}\vec{B}_{Y}+c_{\mathrm{v}}\vec{\omega}+\sigma (\vec{E}_{Y}+\vec{v}\times\vec{B}_{Y}) ,\\&
			\rho_{\mathrm{total}}=\rho_{el}+c_\mathrm{B}\vec{v}.\vec{B}_{Y}+c_{\mathrm{v}}\vec{v}.\vec{\omega}+\sigma_{r}\vec{v}.\vec{E}_Y\\&
			\simeq\rho_{el}=\sum_{i=1}^{n_{G}}\Big[Y_{e_{R}}j^{0}_{{e_i}_R}+N_{w}Y_{e_{L}}j^{0}_{{e_i}_L}+N_{c}\Big(Y_{d_{R}}j^{0}_{{d_i}_R}+Y_{u_{R}}j^{0}_{{u_i}_R}+Y_{Q}N_{w}j^{0}_{Q_{i}}\Big)\Big]+2Y_{H}j^{0}_{H},
		\end{split}
	\end{equation}
where  $j^{0}_{H}$ and $Y_{H}=1$ are the Higgs number density and hypercharge, respectively.

High temperature of the early Universe plasma and low-velocity limit imply that not only $\rho_{\mathrm{total}}\simeq\rho_{el}$ but also the new terms, $T^{\mu\nu}(CME)$ and $T^{\mu\nu}(CVE)$, are small compared to the other terms in the total energy-momentum tensor. Neglecting these terms in the energy-momentum conservation equation (\ref{consT}), we obtain \cite{Abbaslu1}\footnote{The rhs. of Eq.\ (\ref{eq432w}) is actually $\vec{E}_{Y}.\vec{J}$, which is usually neglected.} 
\begin{equation}\label{eq432w}
\partial_{t}\rho+ \vec{\nabla}.\left[(\rho+p)\frac{\vec{v}}{R}\right]+3H(\rho+p)=0%\vec{E}_{Y}.\vec{J},
\end{equation}
\begin{equation}\label{eq4aq}
\begin{split} 
&\left[\partial_{t}\rho+ \frac{1}{R}\vec{\nabla}.\left[(\rho+p)\vec{v}\right]+3H(\rho+p)\right]\vec{v}+\left[\partial_{t}p+H(\rho+p)\right]\vec{v}\\&+(\rho+p)\partial_{t}\vec{v}+(\rho+p)\frac{\vec{v}.\vec{\nabla}}{R}\vec{v}+\frac{\vec{\nabla} p}{R}\\&=
\rho_{\mathrm{total}}\vec{E}_{Y}+\left(\vec{J}\times\vec{B}_{Y}\right).
\end{split} 
\end{equation}

For an incompressible\footnote{The incompressibility condition $\nabla.\vec{v}=0$ leads to $\partial_t \rho+3H(\rho+p)=0$.} and totally hypercharge neutral plasma in the presence of the hypermagnetic field, the momentum equation, Eq.\ (\ref{eq4aq}), reduces to the simple form  \cite{Abbaslu1}
\begin{equation}\label{eq4aq-cv}
\begin{split} 
\partial_{t}\vec{v}=\frac{\vec{J}\times\vec{B}_{Y}}{\rho+p}+\frac{\nu}{{R}^{2}}\nabla^{2}\vec{v},
\end{split} 
\end{equation}
where $\nu\simeq1/(5\alpha_{Y}^{2}T)$ is the kinematic viscosity \cite{{41},{banerjee}}.

The Maxwell's equations may be expressed covariantly as \cite{son1,Yamamoto16,Anand-2017,Landsteiner-2016}
\begin{equation}\label{Maxwell's equatuions}
\begin{split}
&\nabla_{\mu}Y^{\mu\nu}=J^{\nu},\\&
\nabla_{\mu}{\tilde{Y}}^{\mu\nu}=0.
\end{split}
\end{equation}
%
%\textcolor{magenta}{} \textcolor{green}{} \textcolor{green}{, \it i.e.},
%\textcolor{red}{} \textcolor{green}{} \textcolor{green}{,} \textcolor{orange}{()} \textcolor{purple}{}
%
Using Eqs.\ (\ref{Maxwell's equatuions}), and taking the CVE and the CME into account, we obtain the anomalous Maxwell's equations for the hypercharge-neutral plasma in the symmetric phase of the expanding Universe as (see Refs.\ \cite{Giovannini-1997eg,Giovannini-2013oga,{6},{dettmann},Giovannini-2015aea, Abbaslu1})
\begin{equation}\label{eq1}
\frac{1}{R}\vec{\nabla} .\vec{E}_{Y}=0,\qquad\qquad\frac{1}{R}\vec{\nabla}.\vec{B}_{Y}=0,
\end{equation}
\begin{equation}\label{eq2}
\frac{1}{R}\vec{\nabla}\times\vec{ E}_{Y}+\frac{\partial \vec{B}_{Y}}{\partial t}+2H\vec{B}_{Y}=0,
\end{equation}
\begin{equation}\label{eq3}
	\begin{split}
		\frac{1}{R}\vec{\nabla}\times\vec{B}_{Y}&-\left(\frac{\partial \vec{E}_{Y}}{\partial t}+2H\vec{E}_{Y}\right)=\vec{J}\\
		&=\vec{J}_{\mathrm{Ohm}}+\vec{J}_{\mathrm{cv}}+\vec{J}_{\mathrm{cm}},
	\end{split}
\end{equation}
\begin{equation}\label{eq2.3}
	\vec{J}_{\mathrm{Ohm}}=\sigma\left(\vec{E}_{Y}+\vec{v}\times\vec{B}_{Y}\right).
\end{equation}
%\textcolor{red}{
%\begin{equation}\label{eq3.1}
%\vec{J}_{\mathrm{cv}}=c_{\mathrm{v}}\vec{\omega},
%\end{equation}
%\begin{equation}\label{eq3.2}
%\vec{J}_{\mathrm{cm}}=c_{B}\vec{B}_{Y},
%\end{equation}}

In the next section, we identify all of the fast processes which can be considered to be in equilibrium during the evolution process near the EWPT. Then in Sec.\ (\ref{x6}), we use these as conditions to reduce the number of independent variables and the evolution equations.

%\end{itemize}
%%%%%%%%%%%%%%%%%%%%%%%%%%%%%%%%%%%%

\section{Equilibrium conditions}\label{equilibrium}	
In the temperature range of our interest, the Yukawa interactions of all fermions except those of the electron are in thermal equilibrium which leads to the following equilibrium conditions \cite{vh1},
\begin{equation}\label{11tw-1}
\begin{split}
&\mu_{T_{e_i}}:= \mu_{{e_i}_R}-\mu_{{e_i}_L} +\mu_0 =0\quad \quad i=2,3,\\
&\mu_{T_{u_i}}:=\mu_{u_{iR}}-\mu_{Q_{i}}-\mu_0 =0 \quad \quad i=1,2,3,\\
&\mu_{T_{d_i}}:=\mu_{d_{iR}}-\mu_{Q_{i}}+\mu_0 =0\quad \quad i=1,2,3,\\
\end{split}
\end{equation}
where $\mu_0$ is the chemical potential of the Higgs field, and the $\mu_T$'s measure departure from equilibrium. 
Moreover, as mentioned and utilized above, all gauge interactions are in thermal equilibrium. Furthermore, due to the flavor mixing in the quark sector, all up or down quarks belonging to different generations with distinct handedness have the same chemical potential; {\it i.e.,}  $\mu_{u_{iR}}=\mu_{u_R}$, $\mu_{d_{iR}}=\mu_{d_R}$, and $\mu_{Q_i}=\mu_{Q}$  \cite{86}. Therefore, the equilibrium conditions for the quark Yukawa interactions given in Eqs.\ (\ref{11tw-1}) simplify to 
\begin{equation}\label{11tw-2}
\begin{split} 
&\mu_{T_{u}}:=\mu_{u_{R}}-\mu_{Q}-\mu_0 =0,\\
&\mu_{T_{d}}:=\mu_{d_{R}}-\mu_{Q} +\mu_0 =0.
\end{split}
\end{equation}
Combining the above conditions with the hypercharge neutrality condition given in  Eq.\ (\ref{eq59-6}), we obtain
\begin{equation}\label{11tw-6}
6\mu_Q-\mu_{e_L}-\mu_{e_R}-2\mu_{\mu_L}-2\mu_{\tau_L}+13\mu_{0}=0.
\end{equation}
The weak sphaleron processes which try to wash out the asymmetries are in thermal equilibrium, as well \cite{vh1}. Therefore, 
\begin{equation}\label{11tw-3}
\mu_{T_{\mathrm{Sphaleron}}}:= 18\mu_Q+ 2\mu_{e_L}+2\mu_{\mu_L}+2\mu_{\tau_L}=0.
\end{equation}
 There are also  three conservation laws for the three lepton generations, as 
\begin{equation}\label{11tw-5}
\begin{split}
&\frac{\mu_{B}}{3}-\mu_{e_R}-2\mu_{e_L}=0\\&
\frac{\mu_{B}}{3}-\mu_{\mu_R}-2\mu_{\mu_L}=0\\&
\frac{\mu_{B}}{3}-\mu_{\tau_R}-2\mu_{\tau_L}=0.\\&
\end{split}
\end{equation}
%
%\textcolor{magenta}{} \textcolor{green}{} \textcolor{green}{, \it i.e.},
%\textcolor{red}{} \textcolor{green}{} \textcolor{green}{,} \textcolor{orange}{()} \textcolor{purple}{}
%
Using these assumptions, all chemical potentials can be obtained in terms of the chemical potential of the right-handed electron as (see Refs.\ \cite{sh3,Elahi-2020} for more details)
\begin{equation}\label{eq.c3qw}
\begin{split}
&\mu_{e_L} =\mu_{{\nu_e}_L}=-\frac{415 }{962}\mu_{e_R},\quad \quad \mu_{\mu_L}= \mu_{{\nu_\mu}_L}=\frac{59 }{962}\mu_{e_R}, \quad \quad \mu_{\tau_L}= \mu_{{\nu_\tau}_L}=\frac{59 }{962}\mu_{e_R},\\
&\mu_{\mu_R}= \frac{7}{481}\mu_{e_R},\quad\quad \mu_{\tau_R}= \frac{7}{481}\mu_{e_R}, \quad\quad \mu_{0}= \frac{45}{962}\mu_{e_R},\\
& \mu_{B}=12\mu_{Q}= \frac{198}{481}\mu_{e_R}.
\end{split}
\end{equation} 

 Now by using the above equilibrium conditions, the expression for $c_\mathrm{B}$ given in  Eq.\ (\ref{eqc_B}) reduces to,

	\begin{equation}\label{eqc_Bqa} 
	\begin{split}
	c_{B}(t)=& 
	\frac{g'^{2}}{4\pi^{2}}\Big[(\mu_{e_R}-\frac{1}{2}\mu_{e_L})+(\mu_{\mu_R}-\frac{1}{2}\mu_{\mu_L})+(\mu_{\tau_R}-\frac{1}{2}\mu_{\tau_L})+\left(\frac{3}{8}\mu_{B}+3\mu_{0}\right)\Big]\\
	=&\frac{g'^{2}}{4\pi^{2}}\left[(\mu_{T_e}+\mu_{T_\mu}+\mu_{T_\tau}+\mu_{T_d}+4 \mu_{T_u})+\frac{1}{4}\mu_{T_{\mathrm{Sphaleron}}}
		\right].
	\end{split}
	\end{equation}
Upon using Eq.\ (\ref{eq.c3qw}), the expression for $c_\mathrm{B}$ simplifies to 
\begin{eqnarray}\label{simpcB}
c_\mathrm{B}&=& \frac{g'^{2}}{4\pi^{2}}\mu_{T_e}= \frac{g'^{2}}{4\pi^2}\Big(\frac{711}{481}\Big)\mu_{e_R},
\end{eqnarray}

\section{ The evolution equations close to the EWPT} \label{x6}
Now, we shall show how the effective thermal masses can lead to the generation of the hypermagnetic fields and the matter-antimatter asymmetries in the presence of the vorticity.  As stated in Sec.\ \ref{equilibrium}, all chemical potentials of the particles can be obtained in terms of the chemical potential of the right-handed electron. Therefore, it suffices to consider only the coupled evolution equations of the right-hand electron asymmetry, the hypermagnetic field, and the velocity field.

In the following, we choose a simple monochromatic Chern-Simons configuration for the hypermagnetic field $\vec{B}_Y=(1/R)\vec{\nabla} \times \vec{Y}$, and the velocity field $\vec{v}=(1/R)\vec{\nabla} \times \vec{S}$    \cite{{sh2},Abbaslu1,Giovannini-1997eg,Giovannini-2013oga}. To do this, we choose $\vec{Y}=\gamma(t)\left(\sin kz , \cos kz, 0\right)$, and $\vec{S}=r(t)\left(\sin kz , \cos kz, 0\right)$, for their corresponding vector potentials \cite{69,70,76}. 
 Using the aforementioned configurations, the evolution equation for the velocity field, given by Eq.\ (\ref{eq4aq-cv}), reduces to  \cite{Abbaslu1}
	\begin{equation}
		\frac{\partial \vec{v}}{\partial t}=-\nu{k^{\prime}}^{2}\vec{v},
	\end{equation}
	where $k^{\prime}=k/R=kT$. 
	
	Neglecting the displacement current in the lab frame and using the aforementioned configurations, the expression for the hyperelectric field, given by Eq.\ (\ref{eq3}), and the evolution equation for the hypermagnetic field amplitude, given by Eq.\ (\ref{eq2}), reduce to 
\begin{equation}\label{eq16}
		\vec{E}_{Y}=\frac{k^{\prime}}{\sigma }\vec{B}_{Y}-\frac{c_{\mathrm{v}}}{\sigma }k^{\prime}\vec{v}-\frac{c_{B}}{\sigma}\vec{B}_{Y},
\end{equation}
\begin{equation}\label{eq17}
		\begin{split}
			&\frac{dB_{Y}(t)}{dt}=\left[-\frac{1}{ t}-\frac{{k^{\prime}}^{2}}{\sigma } +\frac{c_{B}k^{\prime}}{\sigma } \right]B_{Y}(t)+\frac{c_{\mathrm{v}}}{\sigma}{k^{\prime}}^{2}v(t).
		\end{split}
\end{equation}
Using $\mu_{f}=(6s/T^2)\eta_f$ and Eq.\ (\ref{eq16}), we obtain $\langle\vec{E}_{Y}.\vec{B}_{Y}\rangle$, which appears in the Abelian anomaly equations, as
\begin{equation}\label{eq44}
		\begin{split}
			\langle\vec{E}_{Y}.\vec{B}_{Y}\rangle=\frac{B_{Y}^{2}(t)}{100} \left[\frac{k^{\prime}}{T}-\left(\frac{6sg'^{2}}{4\pi^{2}T^3}\right)\eta_{T_e}
			\right]-\left[\frac{3g'}{256\pi^2}%\textcolor{red}{+\frac{36s^2g'}{8\pi^{2}T^6}\left(\eta_{e_{R}}^{2}-\eta_{e_{L}}^{2}\right)}
			\right]\frac{k^{\prime}T }{100}v(t)B_Y(t).
		\end{split}
\end{equation}
%
%\textcolor{magenta}{} \textcolor{green}{} \textcolor{green}{, \it i.e.},
%\textcolor{red}{} \textcolor{green}{} \textcolor{green}{,} \textcolor{orange}{()} \textcolor{purple}{}
%
The three terms on the right-hand side of Eq.\ (\ref{eq44}), from left to right are due to, the Ampere's law, the CME, and the CVE. Moreover, the term $\eta_{T_e}=\left(\eta_{e_{R}}-\eta_{e_{L}}+\eta_{0}/2\right)=\frac{711}{481}\eta_{e_R}$ originates from the chiral magnetic coefficient as given in Eq.\ (\ref{simpcB}).

Considering the perturbative chirality flip reactions for the right-handed electron, we obtain the evolution equation of its asymmetry $\eta_{e_R}=(n_{e_R}-\bar{n}_{e_R})/s$ as (see Refs.\ \cite{sh1,sh2,28,Abbaslu1,26})
\begin{equation}
\begin{split}
&\frac{d\eta_{{e}_{R}}}{dt}=\frac{g'^{2}}{4\pi^{2} s}\langle\vec{E}_{Y}.\vec{B}_{Y}\rangle+\left(\frac{\Gamma_{0}}{t_\mathrm{EW}}\right)\left(\frac{1-x}{\sqrt{x}}\right)\eta_{T_e},
\end{split}
\end{equation} 
where, $\left(\frac{\Gamma_{0}}{t_\mathrm{EW}}\right)\left(\frac{1-x}{\sqrt{x}}\right)$ is the chirality flip rate, $\Gamma_{0}=121$, $x=\left(t/t_\mathrm{EW}\right)=\left(T_\mathrm{EW}/T\right)^{2}$ is given by the Friedmann law, $t_\mathrm{EW}=\left(M_{0}/2T_\mathrm{EW}^{2}\right)$, $M_{0}=\left(M_\mathrm{Pl}/1.66\sqrt{g^{*}}\right)$, and $M_\mathrm{Pl}$ is the Plank mass.

Using Eq.\ (\ref{eq44}) and the relation $1\mbox{{Gauss}}\simeq2\times10^{-20} \mbox{{GeV}}^{2}$, the evolution equation of the right-handed electron asymmetry and the hypermagnetic field amplitude can be written as
	\begin{equation}\label{eq47z11tw}
\begin{split}
\frac{d\eta_{e_R}}{dx}&=\big[C_{1}-C_{2}\eta_{e_R}(x)
\big]\left(\frac{B_{Y}(x)}{10^{20}G}\right)^{2}x^{3/2}-C_{3}v(x)\left(\frac{B_{Y}(x)}{10^{20}G}\right)\sqrt{x}\\&
-\Gamma_{0}\frac{1-x}{\sqrt{x}}\left(\frac{711}{481}\right)\eta_{e_R}(x),
\end{split}
\end{equation}
\begin{equation}\label{eq49z11tw} 
\begin{split}
\frac{dB_{Y}}{dx}=\frac{1}{\sqrt{x}}\left[-C_{4} +C_{5}\eta_{e_R}(x)
\right]B_{Y}(x)-\frac{1}{x}B_{Y}(x)+C_{6}\frac{v(x)}{x^{3/2}} ,
\end{split}
\end{equation}  
where the coefficients $C_{i}, i=1,...,6$  are given by
\begin{equation}\label{eq51z11tw}
\begin{split}
& C_{1}=9.6\times10^{-4}\left(\frac{k}{10^{-7}}\right)\alpha_{Y},\\&
C_{2}=1.27963\times10^{6} \alpha_{Y}^{2},\\&
C_{3}=2.03719\times10^{-2}\left(\frac{k}{10^{-7}}\right)\alpha_{Y}^{3/2},\\&
C_{4}= 0.356\left(\frac{k}{10^{-7}}\right)^{2},\\&
C_{5}= 4.7061\times10^{8}\left(\frac{k}{10^{-7}}\right)\alpha_{Y},\\&
C_{6}=7.49215\times10^{20}\sqrt{\alpha_{Y}}\left(\frac{k}{10^{-7}}\right)^{2},
\end{split}
\end{equation}
and $\alpha_{Y}=g'^{2}/4\pi\simeq0.01$ is the fine structure constant of $\rm U_Y(1)$. 
%
%\textcolor{magenta}{} \textcolor{green}{} \textcolor{green}{, \it i.e.},
%\textcolor{red}{} \textcolor{green}{} \textcolor{green}{,} \textcolor{orange}{()} \textcolor{purple}{}
%%

For an illustrative example, we consider the velocity as two successive Gaussian pulses with opposite profiles as\footnote{The following expression relates the velocity configuration displayed here to the one introduced at the beginning of this section,
	$$\vec{v}=\frac{kr(x)}{R}\left(\sin kz , \cos kz, 0\right)=v(x)\left(\sin kz , \cos kz, 0\right)$$}
\begin{equation}\label{wqas1}	
v(x)=\frac{v_{0}}{b\sqrt{2\pi}}e^{-\frac{(x-x_{0,1})^{2}}{2b^{2}}}-\frac{v_{0}}{b\sqrt{2\pi}}e^{-\frac{(x-x_{0,2})^{2}}{2b^{2}}}, 
\end{equation}
%	where $b$ denotes the width of the pulses, $v_{0}$ is the amplitude of the velocity fluctuation, and $x_{0,i}$ denotes the times at which fluctuations happen. 
where $b$, $v_{0}$, and $x_{0,i}$ denote the width, the amplitude, and the center time of the fluctuations, respectively. In fact, the occurrence of any fluctuation in a plasma in a quasi-equilibrium state would normally trigger a restoring response originating from dissipative effects, such as viscous effects. Here, for simplicity, we assume that the combined results of the original fluctuations and the ensuing dissipative effects have the Gaussian profiles as given by Eq.\ (\ref{wqas1}) \cite{Temperature fluctuation}.  

\begin{figure*}[]
	%\centering
	\subfigure[]{\label{fig:figure:1cve1}
		\includegraphics[width=.41\textwidth]{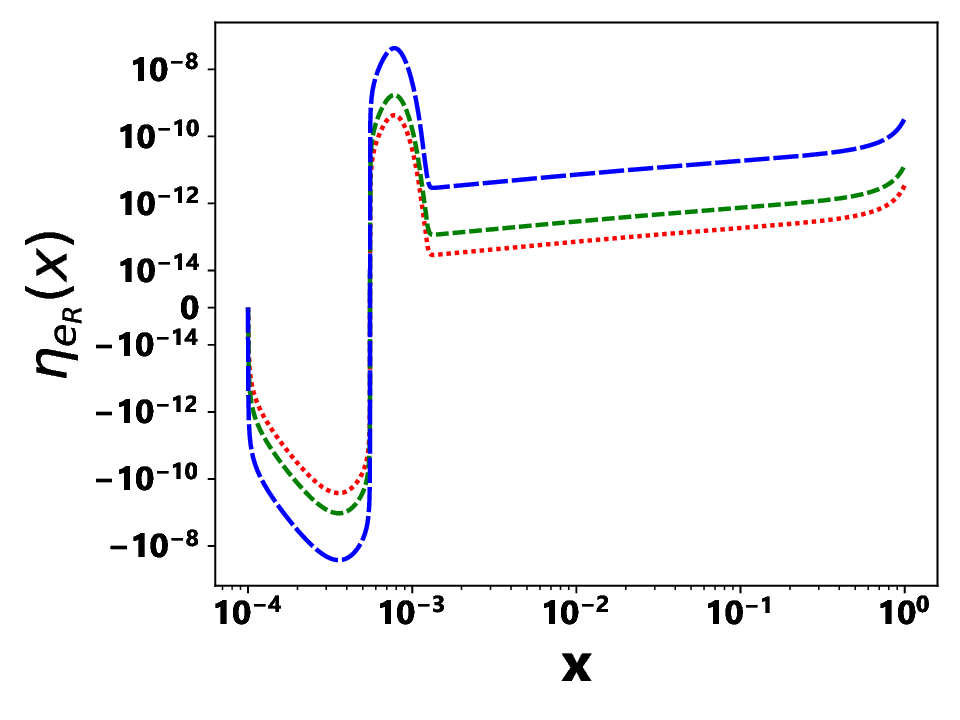}}
	\hspace{8mm}
	\subfigure[]{\label{fig:figure:2cve1}
		\includegraphics[width=.41\textwidth]{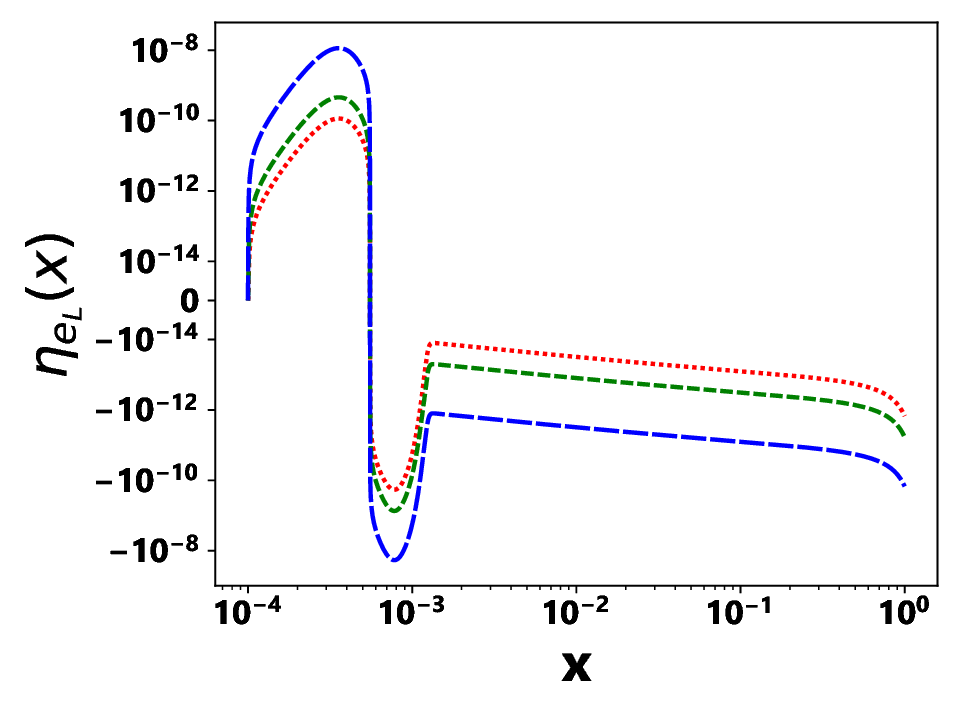}}
	\hspace{8mm}
	\subfigure[]{\label{fig:figure:1cve1r}
		\includegraphics[width=.41\textwidth]{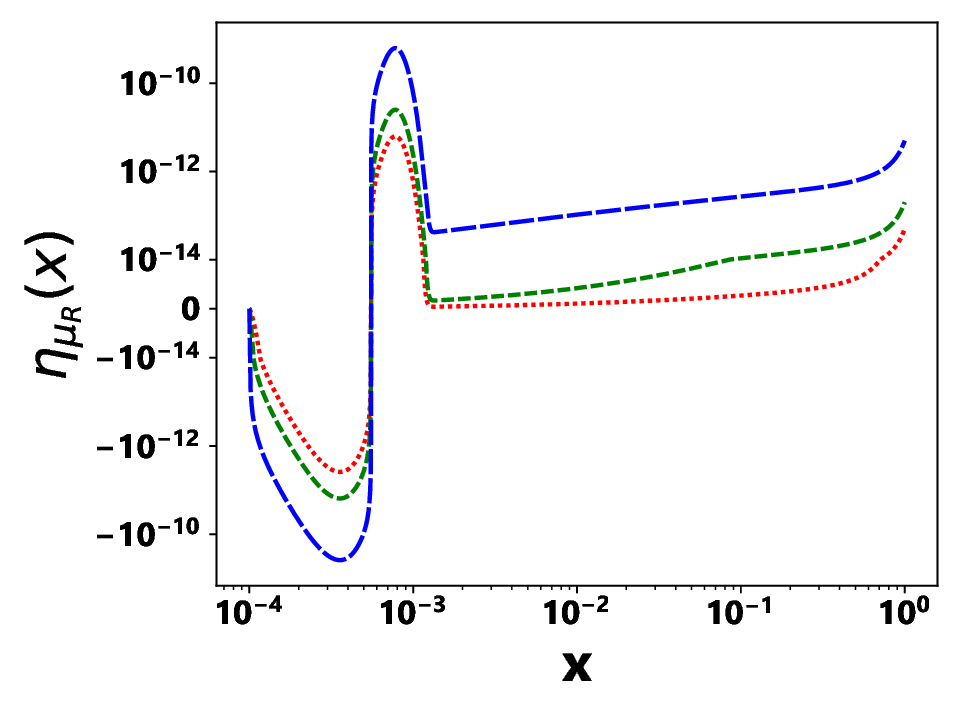}}
	\hspace{8mm}
	\subfigure[]{\label{fig:figure:2cve1r}
		\includegraphics[width=.41\textwidth]{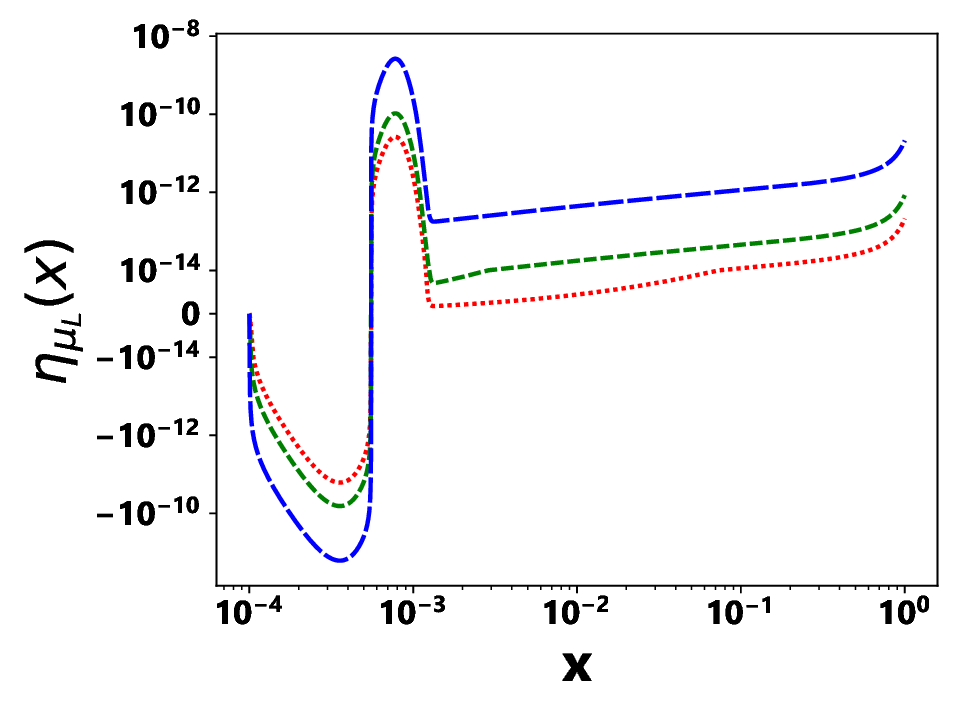}}
	\hspace{8mm}
	\subfigure[]{\label{fig:figure:1cve}
		\includegraphics[width=.41\textwidth]{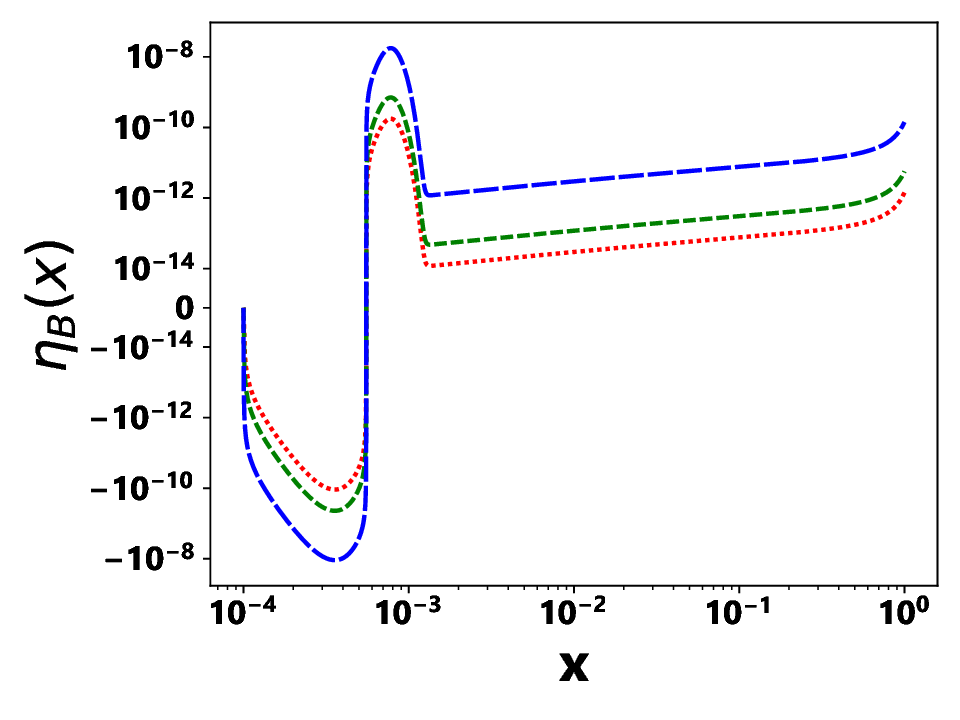}}
	\hspace{20mm}
	\subfigure[]{\label{fig:figure:2cve}
		\includegraphics[width=.41\textwidth]{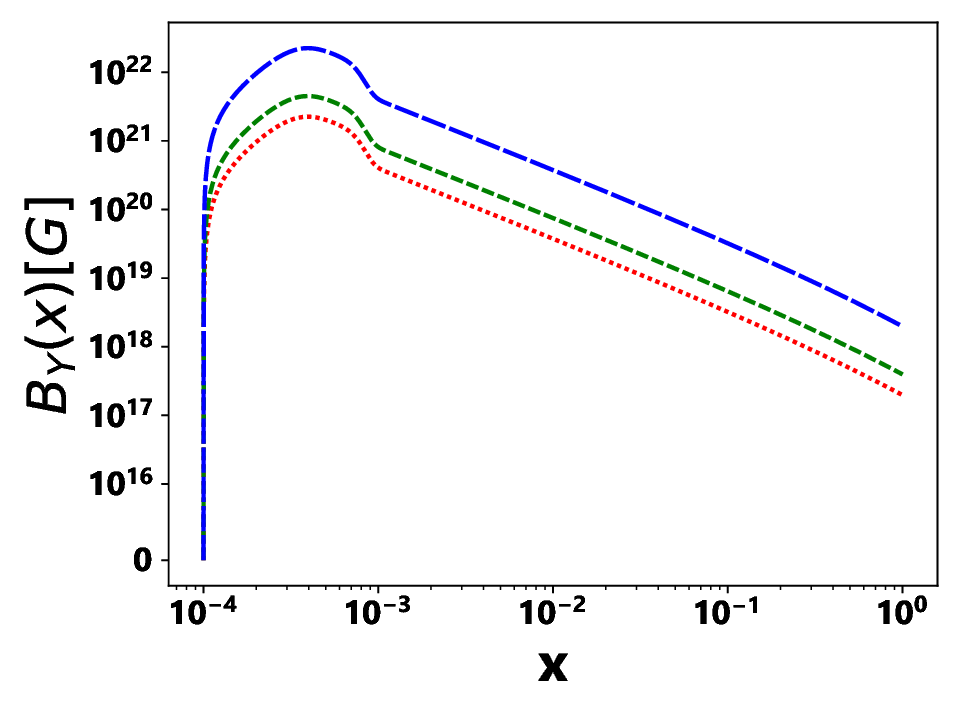}}
	\hspace{8mm}
	\caption{\footnotesize Time plots of the right-handed electron asymmetry $\eta_{e_R}$, the left-handed electron asymmetry $\eta_{e_L}$, the right-handed muon asymmetry $\eta_{\mu_R}$, the left-handed muon asymmetry $\eta_{\mu_L}$, the baryon asymmetry $\eta_{\mathrm{B}}$, and the hypermagnetic field amplitude $B_{Y}$, with the initial conditions $k=10^{-7}$, $B_{Y}^{(0)}=0$, $\eta_{\mathrm{f}}^{(0)}=0$, $b=1\times10^{-4}$, $x_{0,1}=3\times10^{-4}$, and $x_{0,2}=x_{0,1}+5b=8\times10^{-4}$. The dotted-red line is for $v_0=5\times10^{-3}$, the dashed-green line for $v_0=1\times10^{-2}$, and large dashed-blue line for $v_0=5\times10^{-2}$. (According to  Eq.\ (\ref{eq.c3qw}), the asymmetry of the right-handed (left-handed) tau is equal to the asymmetry of the right-handed (left-handed) muon.)  }
	\label{fig4cve}
\end{figure*}

We now set the initial values of the hypermagnetic field amplitude and all matter-antimatter asymmetries, including Higgs asymmetry, to zero, i.e. $B_{Y}^{(0)}=0$,  
$\eta_{R}^{(0)}=0$, and solve the set of evolution equations with the initial conditions $b=1\times10^{-4}$,  $x_{0,1}=3\times10^{-4}$,  $x_{0,2}=x_{0,1}+5b=8\times10^{-4}$, and three different values for $v_0$, and present the results in Fig.\ \ref{fig4cve}. 
As can be seen, upon the occurrence of vorticity fluctuations, the thermal mass induced CVE leads to the generation of strong hypermagnetic fields which then produce the matter-antimatter asymmetries via the Abelian anomaly, all starting from zero initial values, in spite of the fact that the weak sphaleron processes are also taken into account. Figure \ref{fig4cve} shows that by increasing the amplitude of the velocity fluctuations, the maximum and final values of both the hypermagnetic field amplitude and the matter-antimatter asymmetries increase. 
The results also show that the final values of the hypermagnetic field amplitude and the baryon asymmetry at the onset of the EWPT are of the order of $B_{Y}(T_{EW})\sim10^{18}$G and $\eta_{B}(T_{EW})\sim10^{-10}$, respectively. 
To be more specific, for $v_{0}=5\times10^{-2}$, the matter-antimatter asymmetries at $T_{EW}$ are:
$\eta_{B}\approx1.41\times10^{-10}$, $\eta_{e_R}\approx3.44\times10^{-10}$, $\eta_{e_L}=\eta_{\nu_e}\approx-1.48\times10^{-10}$, $\eta_{\mu_R}\approx5.0\times10^{-12}$ and $\eta_{\mu_L}=\eta_{\nu_\mu}\approx 2.11\times10^{-11}$. Moreover, all of the tau asymmetries are identical to those of muon, and the Higgs asymmetry is $\eta_{0}=3.21\times10^{-11}$. One can easily check that theses values satisfy the conditions stated in Sec.~\ref{equilibrium}.

\begin{figure*}[]
	\centering
	\subfigure[]{\label{fig:figure:1cve12}
		\includegraphics[width=.41\textwidth]{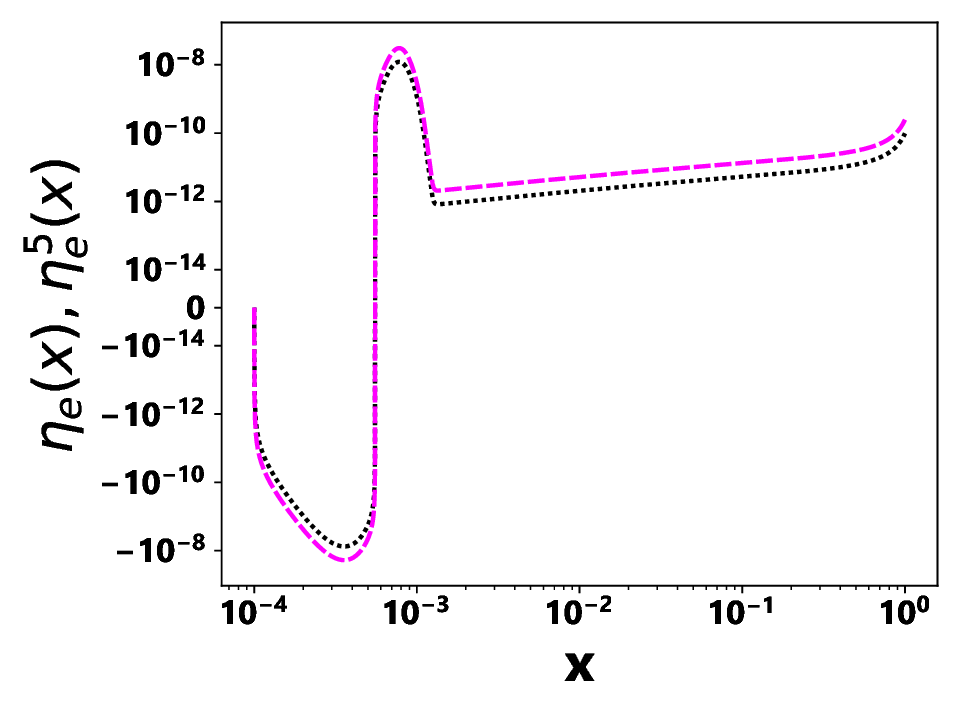}}
	\hspace{8mm}
	\subfigure[]{\label{fig:figure:2cve12}
		\includegraphics[width=.41\textwidth]{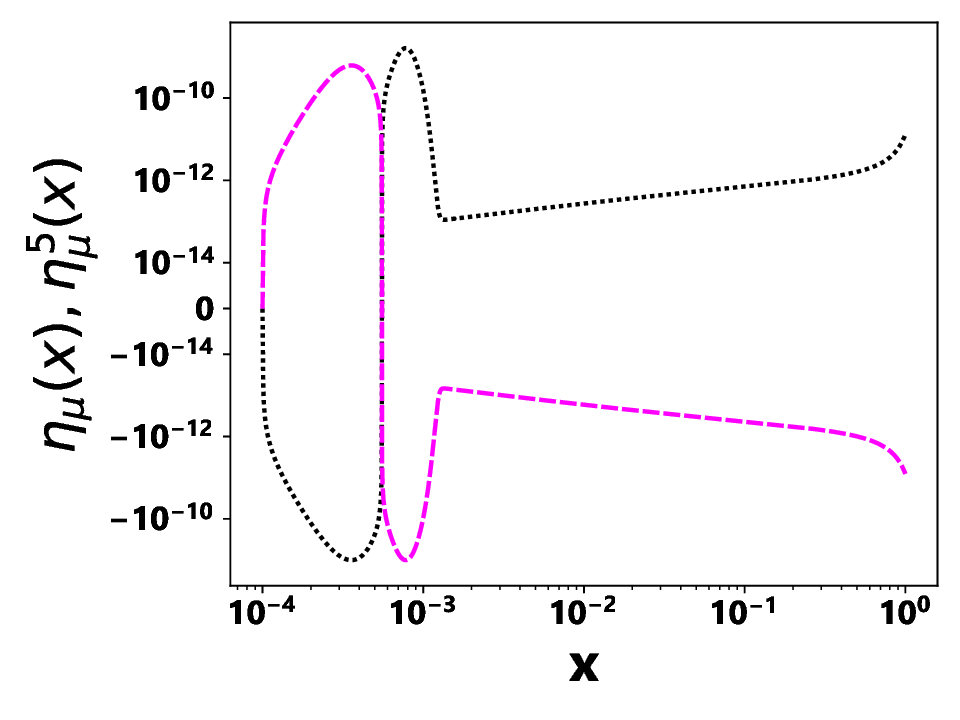}}
	\hspace{8mm}
	\caption{\footnotesize  Time plots of the vector and axial asymmetry of the electron ($\eta_{e},\eta_{e}^{5}$) and muon $\eta_{\mu},\eta_{\mu}^{5}$ with the initial conditions $k=10^{-7}$, $B_{Y}^{(0)}=0$, $\eta_{\mathrm{f}}^{(0)}=0$, $b=1\times10^{-4}$,  $x_{0,1}=3\times10^{-4}$,  $x_{0,2}=x_{0,1}+5b=8\times10^{-4}$, and $v_0=5\times10^{-2}$. The doted line is for the vector asymmetry $\eta$ and the large dashed line is for the axial asymmetry $\eta^5$. The vector/axial asymmetry of the tau is equal to that of the muon, i.e. $\eta_{\tau}=\eta_{\mu}$ and $\eta_{\tau}^{5}=\eta_{\mu}^{5}$. }
	\label{fig4cve22}
\end{figure*}

In Fig.\ (\ref{fig4cve22}), we display the vector and axial vector asymmetries of leptons, for the $v_0=5\times10^{-2}$ case of  Fig.\ (\ref{fig4cve}). As mentioned above, the asymmetries of tau are identical to those of muon. The results show that at the onset of the EWPT, the vector asymmetries of leptons, similar to baryons, are positive. In particular, $\eta_{B}=\eta_{L}\approx1.41\times10^{-10}$. However, while the axial asymmetry of the electron is positive, those of the muon and tau are negative.
The values of these asymmetries play important roles, through the axial anomaly and the CME, in the chiral MHD evolution equations, especially for the evolution of the magnetic field in the broken phase. 
%
%\textcolor{magenta}{} \textcolor{green}{} \textcolor{green}{, \it i.e.},
%\textcolor{red}{} \textcolor{green}{} \textcolor{green}{,} \textcolor{orange}{()} \textcolor{purple}{}
%

%
%\textcolor{magenta}{} \textcolor{green}{} \textcolor{green}{, \it i.e.},
%\textcolor{red}{} \textcolor{green}{} \textcolor{green}{,} \textcolor{orange}{()} \textcolor{purple}{}
%

\newpage
\section{ Conclusion} \label{conclusion}
In this study, we have concentrated on the CVE originating from the temperature-dependent effective thermal masses of the fermions, in the symmetric phase. 
 We have first obtained the hyperelectric CVE coefficient for a single fermion species with chirality $r$ which includes thermal masses, and presented its small thermal mass limit, i.e. $c_{\mathrm{v,}r}(T,m_{r},\mu_r)\simeq rQ_r \left(\frac{T^2}{24}-\frac{m^{2}_{r}(T)}{16\pi^{2}}+\frac{\mu^{2}_{r}(T)}{8\pi^2}\right)$. Adding up the contributions from all fermions in the total hyperelectric chiral vortical current, we have shown that the explicit $T^2$ terms cancel. Moreover, the contributions of the gauge interactions to the thermal masses in the total hyperelectric chiral vortical current cancel out due to the gauge symmetry, while the contributions due to the Yukawa processes yield a nonzero expression proportional to $T^2$. We have emphasized that the explicit $T^2$ dependence in $c_{\mathrm{v,}r}$, which added up to zero in $c_\mathrm{v}$, is distinct from the temperature dependence which entered implicitly from the thermal masses $m_r(T)$.
We have shown that this implicit $T^2$ dependent term can produce a hyperelectric chiral vortical current, in the presence of transient vorticity fluctuation, leading to the generation of hypermagnetic fields and matter-antimatter asymmetries from zero initial values, even though the weak sphalerons are also taken into account.

In this study, we have considered simple monochromatic helical configurations for the vorticity and hypermagnetic fields with positive helicity.
We have investigated the evolution equations of the hypermagnetic field amplitude and the matter-antimatter asymmetry in the early Universe and in the temperature range, $100\mbox{GeV}\le T \le 10\mbox {TeV}$.
 We have shown that, by considering the conservation laws and equilibrium conditions, the asymmetries of all particle species can be expressed in terms of the right-handed electron asymmetry. We have then demonstrated that when small vorticity fluctuations occur relative to the background, the CVE originating from the thermal masses of the fermions becomes active, resulting in the production of the strong hypermagnetic fields, which then produce the matter-antimatter asymmetries from zero initial values, in the presence of the weak sphalerons. This outcome has not been observed in any of the previous studies. Indeed, we have shown that the CVE, which was considered rather inconsequential in the AMHD equations, can have great impact on the generation of hypermagnetic fields and matter-antimatter asymmetries in the early Universe. In fact, we have found that the CME plays a relatively minor role in this model. We have also found that an increase in the amplitudes of the vorticity fluctuations leads to the production of stronger hypermagnetic fields and, therefore, larger matter-antimatter asymmetries. We have also obtained the time evolution of the vector and axial lepton asymmetries up to the EWPT, and observed that for electron $\eta_{e}(T_{EW}),  \eta_{e}^5(T_{EW})>0$,  while for muon and tau $\eta(T_{EW})>0, \eta^5(T_{EW})<0$.
The values of these asymmetries play important roles in the subsequent evolution in the broken phase. 
%
%\textcolor{magenta}{} \textcolor{green}{} \textcolor{green}{, \it i.e.},
%\textcolor{red}{} \textcolor{green}{} \textcolor{green}{,} \textcolor{orange}{()} \textcolor{purple}{}
%

The generated helical hypermagnetic field in this scenario has an amplitude of the order of $B_{Y}(T_ {EW})\sim10^{18}$G. Assuming a sudden transition from symmetric phase to broken phase, this large-scale background field will convert to a large-scale Maxwellian magnetic field through the relation $B_A = B_Y \cos\theta_W\simeq0.88B_Y $. After the EWPT, the evolution of these magnetic fields may be influenced through different processes such as the adiabatic expansion, the axial Abelian anomalous effect, the magnetohydrodynamics turbulent dynamo effect, the viscosity diffusion, and the inverse cascade.

Moreover, the baryon asymmetry produced in this scenario has an amplitude of the order of $\eta_{B}(T_{EW}) \sim10^{-10}$. This asymmetry can change during the electroweak phase transition. However, after the EWPT, the relation $\eta_{B}=(n_{B}-\bar{n}_{B})/(2\pi^{2}g^{*}T^{3}/45)$ and the assumed conservation of the baryon number in the broken phase imply that this asymmetry changes only due to a change in the number of relativistic degrees of freedom $g^{*}$.\\

{\bf Acknowledgments:} S. A. acknowledges the support of the Iran National Science Foundation (INSF) (grant No.\ 4003903).
\begin{appendices}
\section{Small thermal mass limit of the CVE coefficient}\label{Appendix-A}

In this appendix we calculate the small thermal mass limit of the CVE coefficient $\xi_{\mathrm{v,}r}(T,m_r,\mu_r)$ as given by Eq.\ (\ref{eqxiv2}). First, we obtain an equivalent expression as follows:
	\begin{equation}\label{eqxiv-1}
	\begin{split}
	\xi_{\mathrm{v,}r}(T,m_r,\mu_r)&=\frac{r}{8\pi^2}\int_{0}^{\infty}dk\left(\frac{2k^2+m_{r}^2}{E_{k,r}}\right)\tilde{F}_{+}(E_{k,r},\mu_r)\\&=\frac{r}{8\pi^2}\int_{0}^{\infty}dk\left(\frac{k^2}{E_{k,r}}+E_{k,r}\right)\tilde{F}_{+}(E_{k,r},\mu_r)\\&=\frac{r}{8\pi^2}\int_{0}^{\infty}dk\big[\frac{d}{dk}\left(kE_{k,r}\right)\big]\tilde{F}_{+}(E_{k,r},\mu_r)\\&=-\frac{r}{8\pi^2}\int_{0}^{\infty}dk\left(kE_{k,r}\right)\frac{d}{dk}\tilde{F}_{+}(E_{k,r},\mu_r).
	\end{split}
	\end{equation}
Next, we note that any function of $E =
	\sqrt{k^2 + m^2}$ can be expanded about $m^2=0$ as,
	\begin{eqnarray}\label{eqxiv-2}
	f(E)&\simeq& f(E)\Big|_{m^2=0}+\left\{\frac{dE}{dm^2}\frac{dk}{dE}\frac{d}{dk}f[E(k)]\right\}_{m^2=0}m^2 +O(m^4),\nonumber\\
	&\simeq& f(k)+\frac{1}{2k}\frac{d}{dk}f[E(k)]\Big|_{m^2=0}m^2 +O(m^4),
	\end{eqnarray}
where we have used $dE/dm^2=1/(2E)$ and $dk/dE=E/k$. Now we can expand the expression for the CVE coefficient ${\xi_\mathrm{v},r}(T,m_r,\mu_r)$ given in Eq.\ (\ref{eqxiv-1}) to first order in $m^2$ by 
using Eq.\ (\ref{eqxiv-2}) for $\tilde{F}_{+}(E_{k,r},\mu_r)=[e^{\beta(E-\mu_r)}+1]^{-1}+[e^{\beta(E+\mu_r)}+1]^{-1}$ to obtain

%
%\textcolor{magenta}{} \textcolor{green}{} \textcolor{green}{, \it i.e.},
%\textcolor{red}{} \textcolor{green}{} \textcolor{green}{,} \textcolor{orange}{()} \textcolor{purple}{}
%
	\begin{equation}\label{eqxiv-2-t-1}
	\begin{split}
	\xi_{\mathrm{v},r}(T,m_r,\mu_r)\simeq&-\frac{r}{8\pi^2}\int_{0}^{\infty}dk\left(k^2+\frac{m_{r}^2}{2}\right)\frac{d}{dk}\left(\tilde{F}_{+}(k,\mu_r)+\frac{m_{r}^2}{2k}\frac{d\tilde{F}_{+}(E_{k,r},\mu_r)}{dk}\big|_{m^2=0}\right)\\
	\simeq&-\frac{r}{8\pi^2}\int_{0}^{\infty}dk \Big[k^2\left(-\frac{\beta e^{\beta(k-\mu_r)}}{[e^{\beta(k-\mu_r)}+1]^2}-\frac{\beta e^{\beta(k+\mu_r)}}{[e^{\beta(k+\mu_r)}+1]^2}\right)\\&+ \frac{1}{2}km_{r}^2\left(-\frac{\beta^2 e^{\beta(k-\mu_r)}}{[e^{\beta(k-\mu_r)}+1]^2}-\frac{\beta^2 e^{\beta(k+\mu_r)}}{[e^{\beta(k+\mu_r)}+1]^2}\right)\\
	&+\frac{1}{2}km_{r}^2\left(2\frac{\beta^2 e^{2\beta(k-\mu_r)}}{[e^{\beta(k-\mu_r)}+1]^3}+\frac{2\beta^2 e^{2\beta(k+\mu_r)}}{[e^{\beta(k+\mu_r)}+1]^3}\right)\Big].
	\end{split}
	\end{equation}

After a change of variables, the integral can be evaluated to yield,
	\begin{equation}\label{eqxiv-4}
	\begin{split}
	\xi_{\mathrm{v},r}(T,m_r,\mu_r)&\simeq\frac{rT^2}{4\pi^2}\Big[-Li_{2}[-e^{\frac{\mu_{r}}{T}}] -Li_{2}[-e^{-\frac{\mu_{r}}{T}}] -\frac{m_{r}^2}{4T^2}\Big],
	\end{split}
	\end{equation}
	where $Li_{2}[z]$  is the polylogarithm function. Given that in the symmetric phase of the early Universe plasma $\mu_r/T\ll1$, we can use the following series expansion,
	\begin{equation}\label{eqxiv-5}
	\begin{split}
	Li_{2}[-e^{\pm\frac{\mu_{r}}{T}}]\simeq -\frac{\pi^2}{12}\pm\log[2]\frac{\mu_{r}}{T}-\frac{1}{4}(\frac{\mu_{r}}{T})^2\mp\frac{1}{24}(\frac{\mu_{r}}{T})^3+...
	\end{split}
	\end{equation}
	Upon using the above expansions, up to terms of $\mathcal{O}[(\beta\mu)^3]$, in Eq.\ (\ref{eqxiv-4}), the coefficient $\xi_{\mathrm{v},r}(T,m_r,\mu_r)$  for the case $\max(m_r/T,\mu_r/T)\ll1$ takes on the following simple form,
	\begin{equation}\label{xi-1}
	\xi_{\mathrm{v},r}(T,m_{r},\mu_r)\simeq r\big[\frac{T^2}{24}-\frac{m_{r}^2}{16\pi^{2}}+\frac{\mu_{r}^2}{8\pi^2}\big].
	\end{equation}

\section{The numerical values for the CVE coefficient}\label{Appendix-B}

In Table \ref{tablea} of this appendix we display the numerical values for the CVE coefficient $\xi_{\mathrm{v},r}(T,m_r,\mu_r)$ for $\mu=0$, obtained by numerical integration of the exact expression, given in Eq.~(\ref{eqxiv2}) or Eq.\ (\ref{eqxiv-1}) and denoted by $\xi_\mathrm{v,r,N}(T,m_{r},0)$, and the values obtained using its small thermal mass approximation, given in Eq.\ (\ref{eqxiv3}) or Eq.\ (\ref{xi-1}) and denoted by $\xi_\mathrm{v,r,A}(T,m_{r},0)$.
To calculate these quantities, we use the numerical values of thermal masses, also displayed in Table \ref{tablea}. It is interesting to note that although $m_r/T$ is not much less than one, the values of the last two columns are within $0.25\%$, indicating the validity of small thermal mass approximation as expressed in Eq.\ (\ref{eqxiv3}). 
 %\textcolor{orange}{(I think it is better to display the signs, in addition to the absolute values, of $\xi_\mathrm{v,r}(T,m_{r})$, as well. Moreover, the displayed results should have consistent number of significant figures (digits).)}

\begin{table}[!ht]
	\centering 
	\begin{center}
		\begin{tabular}{|p{42mm}|c|c|c|c| c| c|} 
			\hline
			\footnotesize {particle} &$\frac{m_i}{T}$ &  $\frac{\xi_\mathrm{v,r,N}(T,m_i,0)}{rT^2}$ & $\frac{\xi_\mathrm{v,r,A}(T,m_i,0)}{rT^2}$ \\[0.5ex] 
			\hline 
			$e_{R}$, $\mu_R$, $\tau_R$ & $0.1220$ &$0.04157$&$0.04157$ \\
			\hline
			$e_L$, $\mu_L$, $\tau_L$ , $\nu_{e_L}$, $\nu_{\mu_L}$, $\nu_{\tau_L}$&$0.2027$&$0.04140$&$0.04140$ \\
			\hline
			$q_{d_R}$, $q_{s_R}$, $q_{b_R}$, $q_{u_R}$, $q_{c_R}$& $0.4817$&$0.04023$&$0.04019$ \\  
			\hline
			$q_{d_L}$, $q_{s_L}$, $q_{u_L}$, $q_{c_L}$&$0.5178$&$0.04001$&$0.03996$\\
			\hline 
			$q_{b_L}$, $q_{t_L}$&$0.5750$ &$0.03964$&$0.03957$\\  
			\hline
			$q_{t_R}$&$0.5975$ &$0.03948$&$0.03940$\\[1ex] 
			\hline
		\end{tabular}
	\end{center}
	\caption{The effective thermal mass $m_{r}$, the numerical value of the exact expression for the CVE coefficient denoted by $\xi_\mathrm{v,r,N}(T,m_r,0)$, and the numerical value of its small thermal mass approximation $ \xi_\mathrm{v,r,A}(T,m_{r},0)= r\big[\frac{1}{24}-\frac{1}{16\pi^{2}}(\frac{m_{r}}{T})^2\big] T^2$ are displayed for all chiral fermions. We like to point out that the thermal masses have been calculated only to four significant figures, as displayed here.}\label{tablea} 
\end{table}
\newpage
\section{Small thermal mass limit of the CME coefficient}\label{Appendix-C}

To obtain an approximate expression for the chiral magnetic coefficient $\xi_{\mathrm{B},r}(T,m_r,\mu_r)$ for the case $\max(m_r/T,\mu_r/T)\ll1$, we first expand $\tilde{F}_{-}(E_{k,r},\mu_r)$ in  terms of $\beta\mu$ up to $\mathcal{O}[(\beta\mu)^4]$ 
	\begin{equation}\label{q2-wqas}
	\begin{split}
	\xi_{\mathrm{B},r}(T,m_r,\mu_r)&=\frac{rQ_{r}}{4\pi^2}\int_{0}^{\infty}dk\Big[\frac{e^{\beta\mu_r}}{e^{\beta E_{k,r}}+e^{\beta\mu_r}}-\frac{e^{-\beta\mu_r}}{e^{\beta E_{k,r}}+e^{-\beta\mu_r}}\big]
	\\& \simeq\frac{rQ_{r}}{4\pi^2}\int_{0}^{\infty}dk\Big[\frac{\beta ^3 \mu_{r} ^3 e^{\beta  E_{k,r}} \left(-4 e^{\beta E_{k,r}}+e^{2 \beta E_{k,r}}+1\right)}{3 \left(e^{\beta E_{k,r}}+1\right)^4}+\frac{2 \beta  \mu_{r}  e^{\beta E_{k,r}}}{\left(e^{\beta E_{k,r}}+1\right)^2}\Big].
	\end{split}
	\end{equation}
	Now by using $E_{k,r}=\sqrt{k^2+m_{r}^2}$, we expand the above equation in terms of $\beta m$ up to $\mathcal{O}[(\beta m)^4]$. Then the resulting integral can be evaluated exactly as follows,

	\begin{equation}\label{q2-wqsdsd}
	\begin{split}
	\xi_{\mathrm{B},r}(T,m_r,\mu_r)&\simeq\frac{rQ_{r}}{4\pi^2}\int_{0}^{\infty}dk\Big[\left(\frac{2 \beta  \mu_{r}  e^{\beta  k}}{\left(e^{\beta  k}+1\right)^2}+\frac{\beta ^3 \mu_{r} ^3 e^{\beta  k} \left(-4 e^{\beta  k}+e^{2 \beta  k}+1\right)}{3 \left(e^{\beta  k}+1\right)^4}\right)\\&+(\frac{m_{r}}{T})^2 \left(-\frac{\beta \mu  e^{\beta  k} \left(e^{\beta  k}-1\right)}{\beta k \left(e^{\beta  k}+1\right)^3}-\frac{\beta ^3 \mu_{r} ^3 e^{\beta  k} \left(11 e^{\beta  k}-11 e^{2 \beta  k}+e^{3 \beta  k}-1\right)}{6\beta k \left(e^{\beta  k}+1\right)^5}\right)+\mathcal{O}[(\beta m_r)^4]\Big]\\&
	\simeq\frac{rQ_{r}\mu_r}{4\pi^2}\Big[1+0+7\zeta^{'}[-2](\frac{m_{r}}{T})^2+\frac{31}{6} \zeta^{'}[-4](\frac{m_{r}}{T})^2(\frac{\mu_{r}}{T})^2\Big].\\&
	\end{split}
	\end{equation}

%
%\textcolor{magenta}{} \textcolor{green}{} \textcolor{green}{, \it i.e.},
%\textcolor{red}{} \textcolor{green}{} \textcolor{green}{,} \textcolor{orange}{()} \textcolor{purple}{}
%

In Table \ref{tablea1} we compare the numerical values for the CME coefficient $\xi_{\mathrm{B},r}(T,m_r,\mu_r)$, in the limit $\mu_r \rightarrow 0$, obtained by numerical integration of the exact expression, given in Eq.\ (\ref{q2-wqas}) and denoted by $\xi_\mathrm{B,r,N}(T,m_{r},\mu_r)$, and the values obtained using its small thermal mass approximation, given in Eq.\ (\ref{q2-wqsdsd}) and denoted by $\xi_\mathrm{B,r,A}(T,m_{r},\mu_r)$. It is interesting to note that the values of the last two columns are within $0.4\%$, indicating the validity of small thermal mass approximation.

	\begin{table}[!ht]
	\centering 
	\begin{center}
		\begin{tabular}{|p{42mm}|c|c|c|c| c| c|} 
			\hline
			\footnotesize {particle} &$\frac{m_i}{T}$ &  $\frac{\xi_\mathrm{B,r,N}(T,m_i,\mu_r)}{rQ_r\mu_r}$ & $\frac{\xi_\mathrm{B,r,A}(T,m_i,\mu_r)}{rQ_r\mu_r}$ \\[0.5ex] 
			\hline 
			$e_{R}$, $\mu_R$, $\tau_R$ & $0.1220$ &$0.02525$&$0.02525$ \\
			\hline
			$e_L$, $\mu_L$, $\tau_L$ , $\nu_{e_L}$, $\nu_{\mu_L}$, $\nu_{\tau_L}$&$0.2027$&$0.02511$&$0.02511$ \\
			\hline
			$q_{d_R}$, $q_{s_R}$, $q_{b_R}$, $q_{u_R}$, $q_{c_R}$& $0.4817$&$0.02412$&$0.02408$ \\  
			\hline
			$q_{d_L}$, $q_{s_L}$, $q_{u_L}$, $q_{c_L}$&$0.5178$&$0.02394$&$0.02394$\\
			\hline 
			$q_{b_L}$, $q_{t_L}$&$0.5750$ &$0.02363$&$0.02355$\\  
			\hline
			$q_{t_R}$&$0.5975$ &$0.02350$&$0.02340$\\[1ex] 
			\hline
		\end{tabular}
	\end{center}
	\caption{The effective thermal mass $m_{r}$, the numerical value of  $\lim_{\mu \rightarrow 0}[\xi_{\mathrm{B},r}(T,m_r,\mu_r)/(rQ_r\mu_r)]$ based on the exact expression for the CME coefficient given by Eq.\ (\ref{q2-wqas}) and denoted by $\xi_\mathrm{B,r,N}(T,m_r,\mu_r)$, and the numerical value of  $\lim_{\mu \rightarrow 0}[\xi_{\mathrm{B},r}(T,m_r,\mu_r)/(rQ_r\mu_r)]$ based on the small thermal mass approximation for the CME coefficient given by Eq.\ (\ref{q2-wqsdsd}) and denoted by $\xi_\mathrm{B,r,A}(T,m_{r},\mu_r)$, are displayed for all chiral fermions.}\label{tablea1} 
\end{table}
\newpage

\section{Small thermal mass limit of the thermodynamics quantities}\label{Appendix-D}

In this appendix we calculate the small thermal mass limit of the expressions for the number density $n_{i}$, the energy density $\rho_{i}$, the pressure $p_{i}$, and the asymmetry number density $\Delta n_{i}=(n_{i}-\bar{n}_{i})$, for fermion species \lq{\textit{i}}\rq. To obtain an approximate expression for the number density $n_i$ for the case $\max(m_r/T,\mu_r/T)\ll1$, we first expand $f_{FD}(E_{k,r},\mu_r)=[e^{\beta (E_{k,r}-\mu_r)}]^{-1}$ in  terms of $\beta\mu$ up to $\mathcal{O}[(\beta\mu)^3]$  and then expand the resulting expression in terms of $\beta m$ up to $\mathcal{O}[(\beta m)^4]$. Then the resulting integral can be evaluated exactly as follows,
\begin{equation}\label{q2-wqsdgsd}
\begin{split}
n_i(T,m_r,\mu_r)&=\frac{1}{2\pi^2}\int_{0}^{\infty}dk\Big[\frac{k^2}{e^{\beta( E_{k,i}-\mu_i)}+1}\Big]\\&
%\simeq \frac{1}{2\pi^2\beta^3}\int_{0}^{\infty}dk\Big[\frac{k^2}{e^{E_{k,i}}+1}+\frac{\beta\mu_{i}  k^2 e^{E_{k,i}}}{\left(e^{E_{k,i}}+1\right)^2}+\frac{(\beta\mu_{i}) ^2 k^2 e^{ E_{k,i}} \left(e^{E_{k,i}}-1\right)}{2 \left(e^{E_{k,i}}+1\right)^3}+O(\beta^3\mu_{i}^3)\Big]\\&
\simeq  \frac{1}{2\pi^2\beta^3}\int_{0}^{\infty}dk\Big[\frac{k^2}{e^{k}+1}+\frac{\beta\mu_{i}  k^2  e^{k}}{\left(e^{k}+1\right)^2}+\frac{(\beta\mu_{i})^2 k^2 e^{k} \left(e^{k}-1\right)}{2 \left(e^{k}+1\right)^3}\\&-\frac{(\beta m_{i})^2 \left(k e^{k} \left(\beta ^2 \mu_{i}^2-2 \beta \mu_{i} -4 \beta^2 \mu_{i}^2 e^{k}+\beta^2 \mu_{i} ^2 e^{2 k}+2 \beta \mu_{i}  e^{2 k}+4 e^{k}+2 e^{2 k}+2\right)\right)}{4 \left(e^{k}+1\right)^4}\\&+\mathcal{O}[(\beta\mu_i)^3]+\mathcal{O}[(\beta m_i)^4]\Big],\\&
\simeq \frac{T^3}{2\pi^2}\Big[\frac{3 \zeta [3]}{2}+\frac{\pi ^2 }{6}\frac{\mu_{i}}{T}+(\frac{\mu_{i} }{T})^2 \ln[2] +\frac{1}{16} \left(-4 \ln [4]-4(\frac{\mu_{i}}{T}) -(\frac{\mu_{i}}{T})^2 \right)(\frac{m_{i}}{T})^2\Big],
\end{split}
\end{equation}
where $\zeta[3]=1.20206$.
Similarly, we can obtain the small thermal mass limit of the expressions for the energy density $\rho_{i}$, the pressure $p_{i}$, and the asymmetry number density $\Delta n_{i}=(n_{i}-\bar{n}_{i})$ as follows:
\begin{equation}\label{q2-wqsdgsdq}
\begin{split}
\rho_i(T,m_r,\mu_r)&=\frac{1}{2\pi^2}\int_{0}^{\infty}dk\Big[\frac{k^2  E_{k,i}}{e^{\beta( E_{k,i}-\mu_i)}+1}\Big]\\&
%\simeq \frac{1}{2\pi^2\beta^4}\int_{0}^{\infty}dk\Big[\frac{k^2E_{k,i}}{e^{E_{k,i}}+1}+\frac{k^2 \beta\mu_{i}  e^{E_{k,i}} E_{k,i}}{\left(e^{E_{k,i}}+1\right)^2}+\frac{k^2 (\beta\mu_{i}) ^2 e^{E_{k,i}} \left(e^{E_{k,i}}-1\right) E_{k,i}}{2 \left(e^{E_{k,i}}+1\right)^3}+O(\beta^3\mu_{i}^3)\Big]\\&
\simeq \frac{1}{2\pi^2\beta^4}\int_{0}^{\infty}dk\Big[\frac{k^3}{e^{k}+1}+\frac{e^{k} k^3 \beta\mu_{i} }{\left(e^{k}+1\right)^2}+\frac{e^{k} \left(e^{k}-1\right) k^3 (\beta\mu_{i})^2}{2 \left(e^{k}+1\right)^3}
+\Big(+\frac{k}{2 \left(e^{k}+1\right)}-\frac{e^{k} k^2}{2 \left(e^{k}+1\right)^2}\\&+\frac{e^{k} \left(-e^{k} k^2+k^2+e^{k} k+k\right) \beta\mu_{i} }{2 \left(e^{k}+1\right)^3}+\frac{e^{k} \left(4 e^{k} k^2-e^{2 k} k^2-k^2+e^{2 k} k-k\right) (\beta\mu_{i}) ^2}{4 \left(e^{k}+1\right)^4}\Big)(\beta m_{i})^2\\&+\mathcal{O}[(\beta\mu_i)^3]+\mathcal{O}[(\beta m_i)^4]\Big]\\&
\simeq\frac{T^4}{2\pi^2}\Big[\frac{7 \pi ^4}{120}+\frac{9 \beta\mu_{i} \zeta (3)}{2}+\frac{\pi^2 (\beta\mu_{i})^2}{4}+-\frac{1}{24} (\beta m_{i})^2 \left(\pi^2+3\beta\mu_{i} \ln[16]+3 (\beta\mu_{i})^2 \right)\Big],
\end{split}
\end{equation}
\begin{equation}\label{q2-wqsdgwsdq}
\begin{split}
p_i(T,m_r,\mu_r)&=\frac{1}{6\pi^2}\int_{0}^{\infty}dk\Big[\frac{1}{E_{k,i}}\frac{k^4 }{e^{\beta( E_{k,i}-\mu_i)}+1}\Big]\\&
%\simeq\frac{1}{6\pi^2\beta^2}\int_{0}^{\infty}dk\Big[\frac{k^4}{\left(e^{E_{k,i}}+1\right) E_{k,i}}+\frac{k^4 \beta\mu_{i}  e^{E_{k,i}}}{\left(e^{E_{k,i}}+1\right)^2 E_{k,i}}+\frac{k^4 (\beta\mu_{i} )^2 e^{E_{k,i}} \left(e^{E_{k,i}}-1\right)}{2 \left(e^{E_{k,i}}+1\right)^3 E_{k,i}}+O(\beta^3\mu_{i}^3)\Big]\\&
\simeq\frac{1}{6\pi^2\beta^2}\int_{0}^{\infty}dk\Big[+\frac{k^3}{e^{k}+1}+\frac{e^{k} k^3 \beta\mu_{i} }{\left(e^{k}+1\right)^2}+\frac{e^{k} \left(e^{k}-1\right) k^3 \mu_{i} ^2}{2 \left(e^{k}+1\right)^3}+ \Big(-\frac{e^{k} k^2}{2 \left(e^{k}+1\right)^2}-\frac{k}{2 \left(e^{k}+1\right)}\\&
-\frac{e^{k} \left(e^{k} k^2-k^2+e^{k} k+k\right) \beta\mu_{i} }{2 \left(e^{k}+1\right)^3}-\frac{e^{k} \left(-4 e^{k} k^2+e^{2 k} k^2+k^2+e^{2 k}k-k\right) (\beta\mu_{i}) ^2}{4 \left(e^{k}+1\right)^4}\Big)(\beta m_{i})^2\\&+\mathcal{O}[(\beta\mu_i)^3]+\mathcal{O}[(\beta m_i)^4]\Big]\\&
\simeq \frac{T^4}{6\pi^2}\Big[\frac{7\pi^4}{120}+\frac{9\beta\mu_{i}\zeta [3]}{2}+\frac{\pi^2 (\beta\mu_{i})^2}{4}-\frac{1}{8}(\beta m_{i})^2 \left(\pi^2+\beta\mu_{i}\log [16]+3 (\beta\mu_{i})^2  \right)\Big],
\end{split}
\end{equation}
\begin{equation}\label{q2-wqsdgsdqq}
\begin{split}
\Delta n_i(T,m_r,\mu_r)&=\frac{1}{2\pi^2}\int_{0}^{\infty}dk\Big[\frac{k^2}{e^{\beta( E_{k,i}-\mu_{i})}+1}-\frac{k^2}{e^{\beta( E_{k,i}+\mu_{i})}+1}\Big]\\&
%\simeq \frac{1}{2\pi^2\beta^3}\int_{0}^{\infty}dk\Big[\frac{k^2}{e^{E_{k,i}}+1}+\frac{\beta\mu_{i}  k^2 e^{E_{k,i}}}{\left(e^{E_{k,i}}+1\right)^2}+\frac{(\beta\mu_{i}) ^2 k^2 e^{ E_{k,i}} \left(e^{E_{k,i}}-1\right)}{2 \left(e^{E_{k,i}}+1\right)^3}-(\mu_{i}\leftrightarrow-\mu_{i})+O(\beta^3\mu_{i}^3)\Big]\\&
\simeq  \frac{1}{2\pi^2\beta^3}\int_{0}^{\infty}dk\Big[\frac{k^2}{e^{k}+1}+\frac{\beta\mu_{i}  k^2  e^{k}}{\left(e^{k}+1\right)^2}+\frac{(\beta\mu_{i})^2 k^2 e^{k} \left(e^{k}-1\right)}{2 \left(e^{k}+1\right)^3}\\&-\frac{(\beta m_{i})^2 \left(k e^{k} \left(\beta ^2 \mu_{i}^2-2 \beta \mu_{i} -4 \beta^2 \mu_{i}^2 e^{k}+\beta^2 \mu_{i} ^2 e^{2 k}+2 \beta \mu_{i}  e^{2 k}+4 e^{k}+2 e^{2 k}+2\right)\right)}{4 \left(e^{k}+1\right)^4}\\&-(\mu_{i}\leftrightarrow-\mu_{i})+\mathcal{O}[(\beta\mu_i)^3]+\mathcal{O}[(\beta m_i)^4]\Big]\\&
\simeq \frac{T^3}{2\pi^2}\Big[\frac{\pi ^2 }{3}\frac{\mu_{i}}{T} -\frac{1}{2} (\frac{\mu_{i}}{T}) (\frac{m_{i}}{T})^2\Big]\\&
=\frac{\mu_{i}T^2}{6}\Big[1-\frac{3}{2\pi ^2 }  (\frac{m_{i}}{T})^2\Big].
\end{split}
\end{equation}

The  entropy density is related to the pressure and energy density through the relation $s_i=(\rho_{i}+p_{i})/T$.
In Table \ref{table1}, we have displayed the numerical values of the percentage change of the number density $\delta_{n}$, the energy density $\delta_{\rho}$, the pressure $\delta_{p}$, and the asymmetry number density $\delta_{\Delta n}$ for fermion species \lq{\textit{i}}\rq. 
The results show that the thermal masses slightly reduce the values of these thermodynamic quantities in the early Universe plasma. 
	\begin{table*}[!ht]
		\centering 
		\begin{tabular}{|p{37mm}| c|c|c|c| c| c|c|c|c } 
			\hline
			particle  &$\delta_{n}$& $\delta_{\rho}$&$\delta_{p}$&$\delta_{\Delta n}$ \\[0.5ex] 
			\hline 
			$e_{R}$, $\mu_R$, $\tau_R$  &-0.28\%&-0.10\%&-0.32\% &-0.22\% \\
			\hline
			$e_L$, $\mu_L$, $\tau_L$ &-0.78\%&-0.29\%&-0.89\% &-0.62\%  \\
			\hline
			$q_{d_R}$, $q_{s_R}$, $q_{b_R}$, $q_{u_R}$, $q_{c_R}$ &-4.45\% &-1.67\%& -5.03\%&-3.52\% \\  
			\hline
			$q_{d_L}$, $q_{s_L}$, $q_{u_L}$, $q_{c_L}$  &-5.15\%&-1.94\%&-5.82\%&-4.07\% \\
			\hline 
			$q_{b_L}$, $q_{t_L}$  &-6.35\%&-2.39 \%&-7.17\%& -5.02\%\\  
			\hline
			$q_{t_R}$ &-6.86\% &-2.58\% &-7.75\%&-5.42\% \\[1ex] 
			\hline
		\end{tabular}
		\caption{The numerical values of the percentage changes $\delta_{n}$, $\delta_{\rho}$, $\delta_{p}$, and $\delta_{\Delta n}$ due to thermal masses are displayed for all chiral fermions.}\label{table1} 
\end{table*}
%
%\textcolor{magenta}{} \textcolor{green}{} \textcolor{green}{, \it i.e.},
%\textcolor{red}{} \textcolor{green}{} \textcolor{green}{,} \textcolor{orange}{()} \textcolor{purple}{}
%

\end{appendices}

\end{document}